\documentclass[journal,onecolumn]{IEEEtran}
\ifCLASSINFOpdf
\else
\fi
\newcommand{\F}{\mathbb{F}}

\newcommand {\ccc}{{\mathbf{c}}}
\newcommand {\xxx}{{\mathbf{x}}}
\newcommand {\yyy}{{\mathbf{y}}}
\newcommand{\C}{{\mathcal{C}}}

\newcommand{\SSS}{{\mathcal{S}}}

\newcommand{\supp}{{\mathrm{supp}}}
\newcommand{\wt}{{{\rm{wt}}}}

\makeatletter
\newcommand{\rmnum}[1]{\romannumeral #1}
\newcommand{\Rmnum}[1]{\expandafter\@slowromancap\romannumeral #1@}
\makeatother

\usepackage[colorlinks, linkcolor=red, citecolor=blue]{hyperref}
\usepackage{amsmath}
\usepackage{amssymb}
\usepackage{bm}
\usepackage{amssymb}
\usepackage{amsthm}
\usepackage{multirow,booktabs}
\usepackage{cases}
\usepackage{tabularx}
\usepackage{adjustbox}
\usepackage{multirow}
\usepackage[figuresright]{rotating}
\usepackage{longtable}
\usepackage{booktabs}
\usepackage{multirow}
\usepackage{color}
\usepackage{cite}
\usepackage{multirow,booktabs}
\usepackage{cases}
\usepackage{tabularx}
\usepackage{adjustbox}
\usepackage{makecell}
\usepackage[figuresright]{rotating}
\usepackage{longtable}
\usepackage{booktabs}
\usepackage{multirow}
\usepackage{color}
\usepackage{lineno,hyperref,mathtools}
\usepackage{soul}
\usepackage{threeparttable}
\usepackage{booktabs}
\usepackage{pifont}
\usepackage{pgfplots}
\usepackage{amsmath}
\usepackage{float}

\pgfplotsset{compat=1.18}
\allowdisplaybreaks   
\theoremstyle{definition} 
\newtheorem{theorem}{Theorem}

\newtheorem{problem}{Problem}
\newtheorem{lemma}[theorem]{Lemma}

\theoremstyle{definition} 
\newtheorem{remark}{Remark}
\newtheorem{example}{Example}
\newtheorem{definition}{Definition}

\begin{document}

\title{Improved bounds and optimal constructions of pure quantum locally recoverable codes 
\thanks{This research is supported by the Nanyang Technological University Research under Grant 04INS000047C230GRT01.}
}
\author{Yang Li, Shitao Li, Gaojun Luo, San Ling
\thanks{Yang Li is with the School of Physical and Mathematical Sciences, 
Nanyang Technological University, 21 Nanyang Link, Singapore 637371, Singapore
(e-mail: yanglimath@163.com).}
\thanks{Shitao Li is with the School of Internet, Anhui University, Hefei, Anhui 230039, China (e-mail: lishitao0216@163.com).}
\thanks{Gaojun Luo is with the School of Mathematics, Nanjing University of Aeronautics and Astronautics, 
Nanjing, Jiangsu 211106, China (e-mail: gaojun\_luo@nuaa.edu.cn).}
\thanks{San Ling is with the School of Physical and Mathematical Sciences, 
Nanyang Technological University, 21 Nanyang Link, Singapore 637371, Singapore
(e-mail: lingsan@ntu.edu.sg). 
He is also with VinUniversity, Vinhomes Ocean Park, Gia Lam, Hanoi 100000, Vietnam (e-mail: ling.s@vinuni.edu.vn).}
}

\maketitle

\begin{abstract}
By incorporating the concept of locality into quantum information theory, quantum locally recoverable codes (qLRCs) have been proposed, 
motivated by their potential applications in large-scale quantum data storage and their relevance to quantum LDPC codes. 
Despite the progress in optimal quantum error-correcting codes (QECCs), optimal constructions of qLRCs remain largely unexplored, partly due to the fact that 
the existing bounds for qLRCs are not sufficiently tight. 
In this paper, we focus on pure qLRCs derived from the Hermitian construction. 
We provide several new bounds for pure qLRCs and demonstrate that they are tighter than previously known bounds. 
Moreover, we show that a variety of classical QECCs, including quantum Hamming codes, quantum GRM codes, and quantum Solomon-Stiffler codes,  
give rise to pure qLRCs with explicit parameters. Based on these constructions, we further identify 
many infinite families of optimal qLRCs with respect to different bounds, 
achieving code lengths much larger than those of known optimal qLRCs.
\end{abstract}

\section{Introduction}\label{sec.introduction}


Let $\mathbb{C}$ be the {\em field of complex numbers}, and let $\mathbb{C}^q$ be the $q$-dimensional {\em Hilbert space} over $\mathbb{C}$. 
A {\em quantum error-correcting code} (QECC) $\mathcal{Q}$ of length $n$, dimension $\kappa$, minimum distance $\delta$, and alphabet size $q$, 
denoted by $[[n,\kappa,\delta]]_q$, is a $q^{\kappa}$-dimensional subspace of the Hilbert space 
$\mathbb{C}^q\otimes \mathbb{C}^q\otimes\cdots\otimes \mathbb{C}^q:=(\mathbb{C}^q)^{\otimes n}$. 
Such a QECC $\mathcal{Q}$ encodes $\kappa$ logical qudits into entangled states of $n$ physical qudits 
and protects against the erasure of any set of $\delta - 1$ qudits, 
where {\em erasure} in quantum error correction refers to an error whose position in a codeword is known \cite{GBP1997}.

\subsection{Quantum locally recoverable codes}

With the rapid advancement of quantum computing and quantum storage theory \cite{NCV2024,HEHBANS2016}, 
large-scale quantum data storage may become feasible in the future. 
The standard and widely used approach to quantum erasure correction typically relies on measurements \cite{DZ2020}.
However, performing measurements is costly on some physical devices.  
Therefore, it is desirable to design QECCs that can significantly reduce the number of measured qudits \cite{M2025}.

To address this, Golowic and Guruswami \cite{GG-QLRC2025} introduced the concept of quantum locally recoverable codes (qLRCs) with locality 
$r$ by incorporating the notion of locality into QECCs. This framework was later extended to qLRCs with locality 
$(r,\rho)$\footnote{In \cite{GHMM2024}, the notion is termed {\em $(r,\delta)$-locality}. 
In this paper, we instead use $(r,\rho)$-locality to avoid potential confusion with the minimum distance $\delta$ of a QECC.} 
by Galindo $et~al.$ \cite{GHMM2024}. In general, a QECC is said to {\em have locality $(r,\rho)$} 
if, for any erasure at position $i$, there exists a local recovery set $J \ni i$ of size at most $r + \rho - 1$ 
that enables local recovery from any $\rho - 1$ erasures among qudits in $J$, 
and it is said to {\em have locality $r$} if it has locality $(r, 2)$. 
See Definition~\ref{def.qLRCs} for a formal definition.
Furthermore, Golowic and Guruswami also proposed a connection between qLRCs with locality $r$ and 
quantum low-density parity-check (qLDPC) codes, highlighting that such qLRCs serve as a foundational step 
toward studying stronger locality properties in qLDPC codes.

Motivated by their rich applications and theoretical significance, qLRCs with locality $r$ have recently attracted considerable attention.
There are various ways to construct QECCs from classical codes, including the CSS construction and the Hermitian construction \cite{KKKS2006}. 
In the first paper on qLRCs with locality $r$ \cite{GG-QLRC2025}, 
the authors established a useful link between the localities of the QECCs derived from the classical CSS construction 
and those of the classical codes employed. 
Building on this correspondence, Bu {\em et al.} \cite{BGL2025}, Golowic and Guruswami \cite{GG-QLRC2025}, 
Luo {\em et al.} \cite{LCEL2025}, Sharma {\em et al.} \cite{SRT2025}, and Xie {\em et al.} \cite{XZS2025} 
constructed several new families of qLRCs by employing techniques based on 
variations of the hypergraph product, classical Tamo–Barg codes \cite{TB2014}, 
parity-check matrices and cyclic codes, good polynomials, and trace codes, respectively.

For qLRCs derived from the Hermitian construction, a similar connection between the localities of QECCs 
and those of the underlying classical codes was established in \cite{GHMM2024}. 
In the same work, Galindo {\em et al.} proposed a construction of qLRCs with locality 
$(r,\rho)$ based on Hermitian dual-containing maximum distance separable (MDS) codes. 
However, setting $\rho=2$ in their construction does not yield flexible qLRCs with locality $r$.
A detailed discussion of this limitation is given in Remark~\ref{rem.qLRC_from_MDS} in this paper. 
More recently, Li {\em et al.} \cite{LLLLL2025} constructed three families of qLRCs with locality $r$ 
and more flexible parameters from the Hermitian construction by employing near MDS codes that support combinatorial designs.

\subsection{Our motivations}

Similar to classical codes, qLRCs also exhibit various parameter trade-offs. 
Given an $[[n,\kappa,\delta]]_q$ qLRC $\mathcal{Q}$ with locality $r$,
Golowic and Guruswami established a Singleton-like bound in \cite[Theorem 3.5]{GG-QLRC2025}, 
referred to as the {\em GG Singleton-like bound} in this paper, 
as follows: 
\begin{align}\label{eq.GG_Singleton}
        \kappa \leq n-2(\delta-1)-\left\lfloor \frac{n-(\delta-1)}{r+1} \right\rfloor- \left\lfloor \frac{n-2(\delta-1)-\left\lfloor\frac{n-(\delta-1)}{r+1} \right\rfloor}{r+1} \right\rfloor, 
\end{align}
by using quantum mechanical mechanisms. 
In \cite{GG-QLRC2025}, the authors also stated that the GG Singleton-like bound in \eqref{eq.GG_Singleton} applies for arbitrary qLRCs, 
whether pure or impure. Here, an $[[n,\kappa,\delta]]_q$ QECC $\mathcal{Q}$ is {\em pure} if its stabilizer group does not contain 
non-scalar matrices of weight less than $\delta$, and {\em impure} otherwise \cite{KKKS2006}.
Furthermore, a qLRC with parameters meeting the GG Singleton-like bound in \eqref{eq.GG_Singleton} with equality is said to be {\em optimal}.
Generally speaking, it is difficult to determine the purity of a QECC, while many practical QECCs are indeed pure \cite{GLZ2017,KKKS2006}. 
In addition, Luo $et~al.$ \cite{LCEL2025} recently proposed a constraint for pure qLRCs from the CSS construction 
to be optimal with respect to the GG Singleton-like bound in \eqref{eq.GG_Singleton}. 
One can verify that similar situations also hold for pure qLRCs from the Hermitian construction by combining \cite[Theorem 8]{LCEL2025} and \cite[Theorem 6]{LLLLL2025}. 
It is obvious that removing the constraint will bring us more flexibility in constructing optimal pure qLRCs, 
and an effective way is to establish directly some new bounds for pure qLRCs.

On the other hand, the notion of optimality is not restricted to the GG Singleton-like bound in \eqref{eq.GG_Singleton}.
Given any bound for qLRCs and any two parameters selected from $n$, $k$, and $\delta$, if the remaining parameter attains the
{\em largest integer} satisfying the bound (for an {\em upper} bound) or 
the {\em smallest integer} satisfying the bound (for a {\em lower} bound), 
then the corresponding qLRC is said to be optimal with respect to that bound.
In particular, a qLRC whose parameters achieve equality in any other known bounds for qLRCs is optimal.
In this sense, optimal qLRCs are of particular interest, as they either provide the largest error-correcting capability
for a fixed rate or the highest rate for a given error-correcting capability. 
We also say a bound is {\em tight} or {\em useful} if there exist qLRCs with parameters meeting it with equality, 
and we say a bound is {\em tighter than another one} if it yields smaller upper bounds or larger lower bounds on certain parameters than the other one. 
Therefore, given a qLRC with explicit parameters, tighter bounds are helpful to determine its optimality. 
Conversely, for a known bound of qLRCs, it is always interesting to construct an infinite family of qLRCs that meet this bound, 
illustrating this bound is tight or useful. 

Combining the above two aspects, the following two problems naturally arise.

\begin{problem}\label{prob1}
    Can we derive new bounds for pure qLRCs from the Hermitian construction which are tighter than the GG Singleton-like bound in \eqref{eq.GG_Singleton}?
\end{problem}

\begin{problem}\label{prob2}
    If the answer to {\bf Problem} \ref{prob1} is affirmative, can we construct infinite families of optimal pure qLRCs with respect to these new bounds?
\end{problem}

\subsection{Our contributions} 

This paper delves into addressing {\bf Problems} \ref{prob1} and \ref{prob2}. 
We summarize our main contributions as follows. 

\begin{enumerate}
  \item Based on the relationship between pure qLRCs and Hermitian dual-containing classical LRCs (cLRCs),  
  we develop three new bounds for pure qLRCs derived from the Hermitian construction 
  in Theorem \ref{th.pure_bounds}, which are referred to as 
  the pure Griesmer-like bound in \eqref{eq.pure_Griesmer}, 
  the pure Plotkin-like bound in \eqref{eq.pure_Plotkin}, 
  and the pure Sphere-packing-like bound in \eqref{eq.pure_Sphere-packing}, respectively. 
  In particular, the last bound is based on the new Sphere-packing-like bound for cLRCs presented in \eqref{eq.Sphere-packing-like}.
  A detailed discussion in Remark~\ref{rem.sphere-packing} highlights the advantages of our Sphere-packing-like bound over existing bounds.

  \item We demonstrate that these new bounds are tighter than the GG Singleton-like bound in \eqref{eq.GG_Singleton}.
Specifically, for finite code lengths, this result is obtained by reformulating all the bounds in terms of 
the maximum possible dimension $\kappa$ of pure qLRCs with given parameters $n$, $\delta$, and $r$, 
and performing computer-assisted evaluations.
When the code length becomes asymptotically large, we consider the asymptotic forms of these bounds.
Furthermore, we show that the pure Plotkin-like bound in \eqref{eq.Plotkin-like_bound_asymptotic}
is tighter than the pure Griesmer-like bound in \eqref{eq.Griesmer-like_bound_asymptotic},
which in turn is tighter than the pure Singleton-like bound in \eqref{eq.Singleton-like_bound_asymptotic},
and all of these are tighter than the GG Singleton-like bound in \eqref{eq.GG_Singleton-like_bound_asymptotic}. 
We also provide three plots in Figures \ref{fig:bounds_finite}, \ref{fig.1}, and \ref{fig.2} to illustrate these results.
Together with 1) above, these findings give a positive answer to {\bf Problem}~\ref{prob1}. 

  \item For {\bf Problem} \ref{prob2}, we prove that classical QECCs, 
  including quantum Hamming codes, quantum generalized Reed-Muller (GRM) codes, and quantum Solomon-Stiffler codes, 
  can yield pure qLRCs in Theorems \ref{th.Quantum_Hamming} and \ref{th.quantum_GRM} 
  as well as Theorems \ref{th.quantum_SS} and \ref{th.quantum_SS222}, respectively. 
  We collect these pure qLRCs in the following for easy reference. 
  \begin{itemize}
    \item {\bf Quantum Hamming LRCs:}   
    $$\left[\left[\frac{q^{2m}-1}{q^2-1},\frac{q^{2m}-1}{q^2-1}-2m,3\right]\right]_q$$ 
    with locality $r=q^{2m-2}-1$, where $m$ is an integer such that $m\geq 2$ and $\gcd(m,q^2-1)=1$. 

    \item {\bf Quantum GRM LRCs:}
    $$\left[\left[q^{2m},q^{2m}-2\sum_{i=0}^{m}(-1)^i\binom{m}{i}\binom{m+v-iq^2}{v-iq^2}, (s+1)q^{2t}\right]\right]_q$$ 
    with locality $r=(q^2-s+1)q^{2m-2t-2}-1$, 
    where $m$ and $v$ are two integers such that $m\geq 1$ and $0\leq v\leq m(q-1)-1$, 
    and $v+1=t(q^2-1)+s$ with $t\geq 0$ and $0\leq s\leq q^2-2$.

    \item {\bf Quantum Solomon-Stiffer LRCs \Rmnum{1}:} 
    $$\left[\left[\frac{q^{2m}-1}{q^2-1}-\sum_{i=1}^s\frac{q^{2u_i}-1}{q^2-1},\frac{q^{2m}-1}{q^2-1}-\sum_{i=1}^s\frac{q^{2u_i}-1}{q^2-1}-2m,3\right]\right]_q$$
    with locality $r=q^{2m-2}-\sum_{i=1}^sq^{2u_i-2}-1$ for $q>2$, where $m>u_1\geq u_2\geq \cdots\geq u_s\geq 2$ 
    such that $\sum_{i=1}^su_i\leq m$ and at most $q^2-1$ $u_i$'s take the same value, 
    and {\bf Condition A} or {\bf B} in Theorem \ref{th.quantum_SS} is satisfied.

     \item {\bf Quantum Solomon-Stiffer LRCs \Rmnum{2}:} 
    $$\left[\left[\frac{4^{m}-1}{3}-\sum_{i=1}^s\frac{4^{u_i}-1}{3},\frac{4^{m}-1}{3}-\sum_{i=1}^s\frac{4^{u_i}-1}{3}-2m,3\right]\right]_2$$ 
    with locality $r=4^{m-1}-\sum_{i=1}^s4^{u_i-1}-1$, 
    where $m>u_1\geq u_2\geq \cdots\geq u_s\geq 2$, $\sum_{i=1}^s u_i \le m$, and at most three $u_i$’s take the same value.
  \end{itemize}

  \item Based on the constructions presented in 3) above, we further identify many infinite families of optimal pure qLRCs 
  with respect to different bounds. 
  In Table~\ref{tab:qLRCs}, we summarize all known families of optimal pure $[[n,\kappa,\delta]]_{q}$ qLRCs with locality $r$, 
  along with our constructions presented in this paper.
  From the table, it is evident that our optimal constructions are new, as they actually have much larger lengths.  
  Some necessary explanations are also provided below the table.
  These results provide a positive answer to {\bf Problem} \ref{prob2}. 
\end{enumerate}

This paper is structured as follows. After the introduction, 
we present some preliminaries in Section \ref{sec.2} including some new results. 
In Section \ref{sec.3}, we derive three new bounds for pure qLRCs from the Hermitian construction 
and show that they are tighter than the known ones. 
In Section \ref{sec.4}, we employ Hamming codes, GRM codes, and Solomon-Stiffler codes 
to construct pure qLRCs with explicit parameters. 
Based on these constructions, we further identify many infinite families of optimal pure qLRCs with respect to different bounds. 
Finally, we conclude the paper in Section \ref{sec.conclusion}.

\section{Preliminaries}\label{sec.2}

Let $\F_q$ denote the {\em finite field} of size $q$, and let $\F_q^*=\F_q\setminus \{0\}$ denote its {\em multiplicative group}, 
where $q=p^h$ is a prime power. 
An $[n,k,d]_q$ {\em linear code} $\C$ is a $k$-dimensional linear subspace of $\F_q^n$ with minimum distance $d:=d(\C)$. 
Given an $[n,k]_{q^2}$ linear code $\C$, 
its {\em Hermitian dual} is an $[n,n-k]_{q^2}$ linear code defined by 
\begin{align*}
    \C^{\perp_{\rm H}}:=\left\{\yyy=(y_1,y_2,\ldots,y_n)\in \F_{q^2}^n:~\sum_{i=1}^nx_iy_i^q=0,~\forall~ \xxx=(x_1,x_2,\ldots,x_n)\in \C\right\}.
\end{align*}
Note that removing the exponent $q$ in the definition of $\C^{\perp_{\rm H}}$ gives the definition of the {\em (Euclidean) dual} $\C^{\perp}$. 
Write $\C^q:=\{\ccc^q=(c_1^q,c_2^q,\ldots,c_n^q): \ccc\in \C\}$. 
Then it is easy to check that $\C^{\perp_{\rm H}}=(\C^q)^{\perp}=(\C^{\perp})^q$  
and $d^{\perp_{\rm H}}:=d(\C^{\perp_{\rm H}})=d(\C^{\perp})$. 
A $q^2$-ary linear code $\C$ is said to be {\em Hermitian self-orthogonal} if $\C\subseteq \C^{\perp_{\rm H}}$, 
and {\em Hermitian dual-containing} if $\C^{\perp_{\rm H}}\subseteq \C$. 
Moreover, $\C$ is Hermitian self-orthogonal if and only if $\C^{\perp_{\rm H}}$ is Hermitian dual-containing.

\subsection{Classical locally recoverable codes}

We first recall some basic results on classical locally recoverable codes (cLRCs). 
Given a codeword $\ccc=(c_1,c_2,\ldots,c_n)\in \C$, 
define its {\em support} as $\supp(\ccc)=\{1\leq i\leq n:~c_i\neq 0\}$. 
Then a cLRC can be mathematically defined as follows. 

\begin{definition}{\em (\!\! \cite[Definition 1]{LCEL2025})}\label{def.cLRC}
  An $[n,k,d]_q$ linear code $\C$ with dual $\C^{\perp}$ is said to be a {\em classical locally recoverable code (cLRC) with locality $r$} 
  if, for each $i\in \{1,2,\ldots,n\}$, 
  there exists some codeword $\ccc^{\perp}=(c_1,c_2,\ldots,c_n)\in \C^{\perp}$ 
  such that $i\in \supp(\ccc^{\perp})$ and $|\supp(\ccc^{\perp})|\leq r+1$. 
  Moreover, we abbreviate such a code $\C$ as an {$(n,k,d,q;r)$ cLRC}, or simply an {$r$-cLRC}. 
\end{definition} 

Given an $[n,k,d]_{q^2}$ linear code $\C$, 
based on Definition \ref{def.cLRC} and the relation $\C^{\perp_{\rm H}}=(\C^q)^{\perp}$, 
we can directly conclude that $\C^{\perp}$ and $\C^{\perp_{\rm H}}$ have the same locality.
Similar to classical linear codes, cLRCs exhibit various parameter trade-offs.
Several fundamental bounds on cLRCs have been established in the literature using tools from coding theory and combinatorics. 
Given an $(n,k,d,q;r)$ cLRC $\C$, 
its parameters satisfy the {\em Singleton-like bound} \cite{GHSY2012}: 
\begin{align}\label{eq.cLRC_Singleton_bounds}
    d\leq n-k-\left \lceil \frac{k}{r} \right \rceil+2. 
\end{align}
Since the Singleton-like bound in \eqref{eq.cLRC_Singleton_bounds} is independent of the field size $q$, 
it is not tight for small $q$ and large $n$. 
In \cite{CM2013}, Cadambe and Mazumdar established an alphabet-dependent bound for cLRCs, 
namely the {\em CM bound}:
\begin{align}\label{eq.cLRC_CM_bounds}
    k\leq \min_{0\leq \ell\leq \left\lceil \frac{n-1}{r+1}\right\rceil} \left\{ \ell r+k_{\rm opt}^{(q)}(n-\ell (r+1), d) \right\},
\end{align}
where $k_{\rm opt}^{(q)}(n,d)$ is the maximum dimension of a $q$-ary linear code with length $n$ and minimum distance $d$. 

However, the value of $k_{\mathrm{opt}}^{(q)}(n,d)$ is not always easy to obtain,
which makes the CM bound in \eqref{eq.cLRC_CM_bounds} non-explicit.
In particular, by applying the Griesmer and Plotkin bounds \cite{HKS2021} for linear codes in \eqref{eq.cLRC_CM_bounds},
Hao $et~al.$~\cite{HXSCFY2020} derived two more explicit bounds for cLRCs, 
namely the {\em Griesmer-like bound} and the {\em Plotkin-like bound}.
These two bounds were originally proposed for integers $\ell$ satisfying $1 \le \ell \le \left\lceil \frac{k}{r} \right\rceil - 1$.
Nevertheless, they can be naturally extended to the case $\ell = 0$ by including the classical Griesmer and Plotkin bounds,
as shown below, where we also present a new {\em Sphere-packing-like bound} via a similar method.

\begin{lemma}
	Let $\C$ be an $(n,k,d,q;r)$ cLRC and let $\ell$ be an integer.
    Then the following bounds hold. 
    \begin{itemize}
        \item [\rm 1)] {\em (\!\! Griesmer-like bound \cite{HXSCFY2020,HKS2021})} We have that   
        \begin{align}\label{eq.Griesmer-like}
            n\geq \max_{0\leq \ell\leq \left\lceil \frac{k}{r} \right\rceil-1}
                                                                \left\{\ell(r+1)+\sum_{i=0}^{k-\ell r-1}\left\lceil \frac{d}{q^i} \right\rceil \right\}.
        \end{align}

        \item [\rm 2)] {\em (\!\! Plotkin-like bound \cite{HXSCFY2020,HKS2021})} We have that  
        \begin{align}\label{eq.Plotkin-like}
            d\leq \min_{0\leq \ell\leq \left\lceil \frac{k}{r} \right\rceil-1}
                                                                \left\{\frac{q^{k-\ell r-1}(q-1)(n-\ell(r+1))}{q^{k-\ell r}-1} \right\}.
        \end{align}

        \item [\rm 3)] {\em (\!\! Sphere-packing-like bound)} We have that   
        \begin{align}\label{eq.Sphere-packing-like}
           k\leq n-\max_{0\leq \ell\leq \left\lfloor \frac{n-1}{r+1}\right\rfloor} \left\{ \ell +
                \log_q \left(\sum_{i=0}^{\lfloor \frac{d-1}{2} \rfloor} \binom{n-\ell(r+1)}{i} (q-1)^i \right) \right\}.
        \end{align}
    \end{itemize}
\end{lemma}
\begin{IEEEproof}
    1) and 2) have been reported in \cite{HXSCFY2020} and \cite{HKS2021}. 
    For 3), it follows from using directly the classical Sphere-packing bound \cite[Theorem 1.9.6]{HKS2021} in \eqref{eq.cLRC_CM_bounds} that 
      \begin{align*}
            k & \leq \min_{0\leq \ell\leq \left\lfloor \frac{n-1}{r+1}\right\rfloor} \left\{ \ell r+ 
            \log_q \left(\frac{q^{n-\ell(r+1)}} {\sum_{i=0}^{\lfloor \frac{d-1}{2} \rfloor} \binom{n-\ell(r+1)}{i} (q-1)^i} \right) \right\} \\ 
              & = \min_{0\leq \ell\leq \left\lfloor \frac{n-1}{r+1}\right\rfloor} \left\{ \ell r+ 
            n-\ell(r+1) - \log_q \left(\sum_{i=0}^{\lfloor \frac{d-1}{2} \rfloor} \binom{n-\ell(r+1)}{i} (q-1)^i \right) \right\} \\ 
              & = n-\max_{0\leq \ell\leq \left\lfloor \frac{n-1}{r+1}\right\rfloor} \left\{ \ell +
                \log_q \left(\sum_{i=0}^{\lfloor \frac{d-1}{2} \rfloor} \binom{n-\ell(r+1)}{i} (q-1)^i \right) \right\}.
      \end{align*}
\end{IEEEproof}

\begin{remark}\label{rem.sphere-packing}
There also exist several other forms of Sphere-packing-like bounds for cLRCs.
In \cite[Corollary~3]{ABHMT2018}, Agarwal $et~al.$ presented a Sphere-packing-like bound for $(r,\rho)$-cLRCs.
Although this bound can be directly applied to $r$-cLRCs by setting $\rho=2$, it was shown to be slightly weaker than 
the Singleton-like bound in \eqref{eq.cLRC_Singleton_bounds} (see \cite[Page~4168]{WZL2019}).
Furthermore, by investigating the Sphere-packing problem in a specially defined space for cLRCs, 
Wang $et~al.$ established three Sphere-packing-like bounds in \cite[Theorem~11]{WZL2019} (Bound~$A^{+}$), 
\cite[Theorem~13]{WZL2019} (Bound~$B^{+}$), and \cite[Theorem~7]{WZL2019} (Bound~$C$), respectively.
However, Bound~$A^{+}$ only applies to cLRCs with disjoint local repair groups, 
Bound~$B^{+}$ is restricted to cLRCs with minimum distance $d \ge 5$ and locality $2 \le r \le \frac{n}{2} - 2$, 
and Bound~$C$ is valid only for binary cLRCs with locality $r=2$.
Compared with these four bounds, our Sphere-packing-like bound in \eqref{eq.Sphere-packing-like} 
eliminates all these constraints, making it more general and easier to use.
\end{remark}

\subsection{Quantum locally recoverable codes}
Following the notation used in \cite{GHMM2024}, we recall the definition of quantum locally recoverable codes (qLRCs) as follows.

\begin{definition}{\em (\!\! \cite[Definitions 9 and 10]{GHMM2024})}\label{def.qLRCs}
  A QECC $\mathcal{Q}\subseteq (\mathbb{C}^q)^{\otimes n}$ is said to be a 
  {\em quantum locally recoverable code (qLRC) with locality $r$}, simply an $r$-qLRC if, 
  for each $i\in \{1,2,\ldots,n\}$, there exists a set $J\subseteq \{1,2,\ldots,n\}$ containing $i$ with $|J|\leq r+1$ 
  such that for every subset $I\subsetneq J$ with $|I|=1$, there exists a trace-preserving map $\mathcal{R}_{\mathcal{Q},I}^J$, 
  which acts only on the qudits corresponding to $J$ and keeps untouched the remaining ones, such that 
  $$
  \mathcal{R}_{\mathcal{Q},I}^J \circ \Gamma^I(|\varphi\rangle \langle \varphi|) = |\varphi\rangle \langle \varphi|
  $$
  for any $|\varphi\rangle \in \mathcal{Q}$, where $\Gamma^I$ is a mapping given as in \cite[Page 6]{GHMM2024}.
\end{definition}

There are various types of QECCs, and the so-called {\em quantum stabilizer codes} form the most-known class. 
Quantum stabilizer codes have close connections to classical codes, 
containing those derived from the CSS construction, Hermitian construction, and symplectic construction \cite{KKKS2006}. 
In particular, for $\mathcal{Q}$ being a stabilizer code, Definition \ref{def.qLRCs} can be reduced to an easier form 
by considering cLRCs with special structures. 
The following lemma gives the Hermitian construction for qLRCs.


\begin{lemma}\label{lem.Hermitian_qLRC}
{\em (\!\! Hermitian construction for qLRCs \cite[Theorem 29]{GHMM2024})} 
  If $\C$ is an $[n,k,d]_{q^2}$ Hermitian dual-containing code with locality $r$ and $d^{\perp_{\rm H}}\geq 2$, 
then there exists an 
$$\left[\left[n,\kappa,\delta\right]\right]_q$$ 
qLRC $\mathcal{Q}$ with locality $r$, 
where 
\begin{align*}
    \kappa = 2k-n, \quad \delta = \left\{
    \begin{array}{ll} 
        \wt(\C\setminus \C^{\perp_{\rm H}})\geq d, & \mbox{if}~ \C^{\perp_{\rm H}}\subsetneq \C, \\
        d, & \mbox{if}~ \C^{\perp_{\rm H}}= \C, 
    \end{array}
        \right.
\end{align*}
and $\wt(\SSS)$ denotes the minimum weight of all nonzero vectors in any nonempty set $\SSS\subseteq \F_{q^2}^n$.
Furthermore, such a qLRC $\mathcal{Q}$ is said to be {\em pure} if $\delta = d$; and {\em impure} otherwise.
\end{lemma}

\section{Improved bounds of pure qLRCs}\label{sec.3}

This section focuses on {\bf Problem} \ref{prob1}. 
We establish some new and explicit bounds for pure qLRCs and show that they are tighter than 
the known ones.  
First of all, as confirmed in \cite[Theorem 31 and Remark 34]{GHMM2024} and \cite[Remark 2]{LLLLL2025}, a pure $[[n,\kappa,\delta]]_q$ qLRC with locality $r$ derived 
from the Hermitian construction satisfies the {\em pure Singleton-like bound}  
        \begin{align}\label{eq.pure_Singleton}
            2\delta \leq n - \kappa - 2\left\lceil \frac{n+\kappa}{2r} \right\rceil + 4.
        \end{align}
The following theorem provides another three explicit bounds for pure qLRCs. 

\begin{theorem}\label{th.pure_bounds}
    Let $\mathcal{Q}$ be a pure $[[n,\kappa,\delta]]_q$ qLRC with locality $r$ constructed by the Hermitian construction 
    given in Lemma \ref{lem.Hermitian_qLRC}. 
    Then the following bounds hold. 
    \begin{enumerate}
        \item {\em (Pure Griesmer-like bound)}
        We have that  
        \begin{align}\label{eq.pure_Griesmer}
            n \geq \max_{0\leq \ell\leq \left\lceil \frac{n+\kappa}{2r} \right\rceil-1}
                \left\{\ell(r+1)+\sum_{t=0,~2\mid t}^{n+\kappa-2\ell r-2} \left\lceil \frac{\delta}{q^{t}} \right\rceil \right\}.
        \end{align}

        \item {\em (Pure Plotkin-like bound)}
        We have that 
        \begin{align}\label{eq.pure_Plotkin}
            \delta \leq \min_{0\leq \ell\leq \left\lceil \frac{n+\kappa}{2r} \right\rceil - 1}
                \left\{ \frac{q^{n+\kappa - 2\ell r - 2}(q^2 - 1)(n - \ell(r+1))}{q^{n+\kappa - 2\ell r} - 1} \right\}.
        \end{align}

        \item {\em (Pure Sphere-packing-like bound)}
        We have that 
        \begin{align}\label{eq.pure_Sphere-packing}
            \kappa \leq n- 2 \max_{0\leq \ell\leq \left\lfloor \frac{n-1}{r+1}\right\rfloor} \left\{\ell+ \log_{q^2}\left(
                \sum_{i=0}^{\lfloor \frac{\delta-1}{2} \rfloor} \binom{n-\ell(r+1)}{i}\left(q^2-1\right)^i \right) \right\}.
        \end{align}
    \end{enumerate}
\end{theorem}
\begin{IEEEproof}
    Since $\mathcal{Q}$ is a pure $[[n,\kappa,\delta]]_q$ qLRC with locality $r$ constructed by the Hermitian construction 
    given in Lemma \ref{lem.Hermitian_qLRC}, then there exists a corresponding Hermitian dual-containing code $\C$ with 
    parameters 
    \begin{align}\label{eq.pure_parameters}
        \left(n,\frac{n+\kappa}{2},\delta,q^2;r\right)  
    \end{align}
    and $d^{\perp_{\rm H}}\geq 2$. 
    {Since the locality of a linear code is naturally no larger than its dimension, we have $r\leq \frac{n+\kappa}{2}$ and $\left \lceil \frac{n+\kappa}{2r}\right\rceil\geq 1$}. 
    Applying the Griesmer-like bound in \eqref{eq.Griesmer-like} to $\C$ with parameters given in \eqref{eq.pure_parameters}, 
    we have  
    \begin{align}\label{eq.qLRC_bound_111}
        n\geq \max_{0\leq \ell\leq \left\lceil \frac{n+\kappa}{2r} \right\rceil-1}
                                                                \left\{\ell(r+1)+\sum_{i=0}^{\frac{n+\kappa}{2}-\ell r-1}\left\lceil \frac{\delta}{q^{2i}} \right\rceil \right\}.
    \end{align}
    Taking $t=2i$ in \eqref{eq.qLRC_bound_111} immediately yields the result described in \eqref{eq.pure_Griesmer}.
    Similarly, applying the Plotkin-like bound in \eqref{eq.Plotkin-like} and 
    the Sphere-packing-like bound in \eqref{eq.Sphere-packing-like} to $\C$ yields 
    \eqref{eq.pure_Plotkin} and \eqref{eq.pure_Sphere-packing}, respectively. 
    This completes the proof.
\end{IEEEproof}

The following remark establishes a connection between optimal cLRCs and pure qLRCs. 

\begin{remark}
    Let $\C$ be an $(n,k,d,q^2;r)$ Hermitian dual-containing cLRC and let $\mathcal{Q}$ be the corresponding pure $[[n,2k-n,d]]_q$ qLRC with locality $r$ 
    constructed by the Hermitian construction given in Lemma \ref{lem.Hermitian_qLRC}. 
    From the proof of Theorem \ref{th.pure_bounds}, it is easy to see that
    $\mathcal{Q}$ is optimal with respect to a certain pure qLRC bound in 
    \eqref{eq.pure_Singleton}, \eqref{eq.pure_Griesmer}, \eqref{eq.pure_Plotkin}, or \eqref{eq.pure_Sphere-packing} 
    if and only if 
    $\C$ is optimal with respect to the corresponding cLRC bound in 
    \eqref{eq.cLRC_Singleton_bounds}, \eqref{eq.Griesmer-like}, \eqref{eq.Plotkin-like}, or \eqref{eq.Sphere-packing-like}, respectively. 
\end{remark}


Now, we have four explicit bounds for pure qLRCs derived from the Hermitian construction in Lemma \ref{lem.Hermitian_qLRC}. 
In order to evaluate their usefulness, we need to compare these bounds to see which is the tightest one and 
whether they are tighter than the GG Singleton-like bound in \eqref{eq.GG_Singleton}. 
Since these bounds are established for different sets of parameters, they cannot be directly compared. 
Depending on whether the code lengths are finite or asymptotically large, 
we divide our comparison into two parts, performing the corresponding variations for each case.

\begin{figure}[htbp]  
    \centering
    \includegraphics[width=0.85\textwidth]{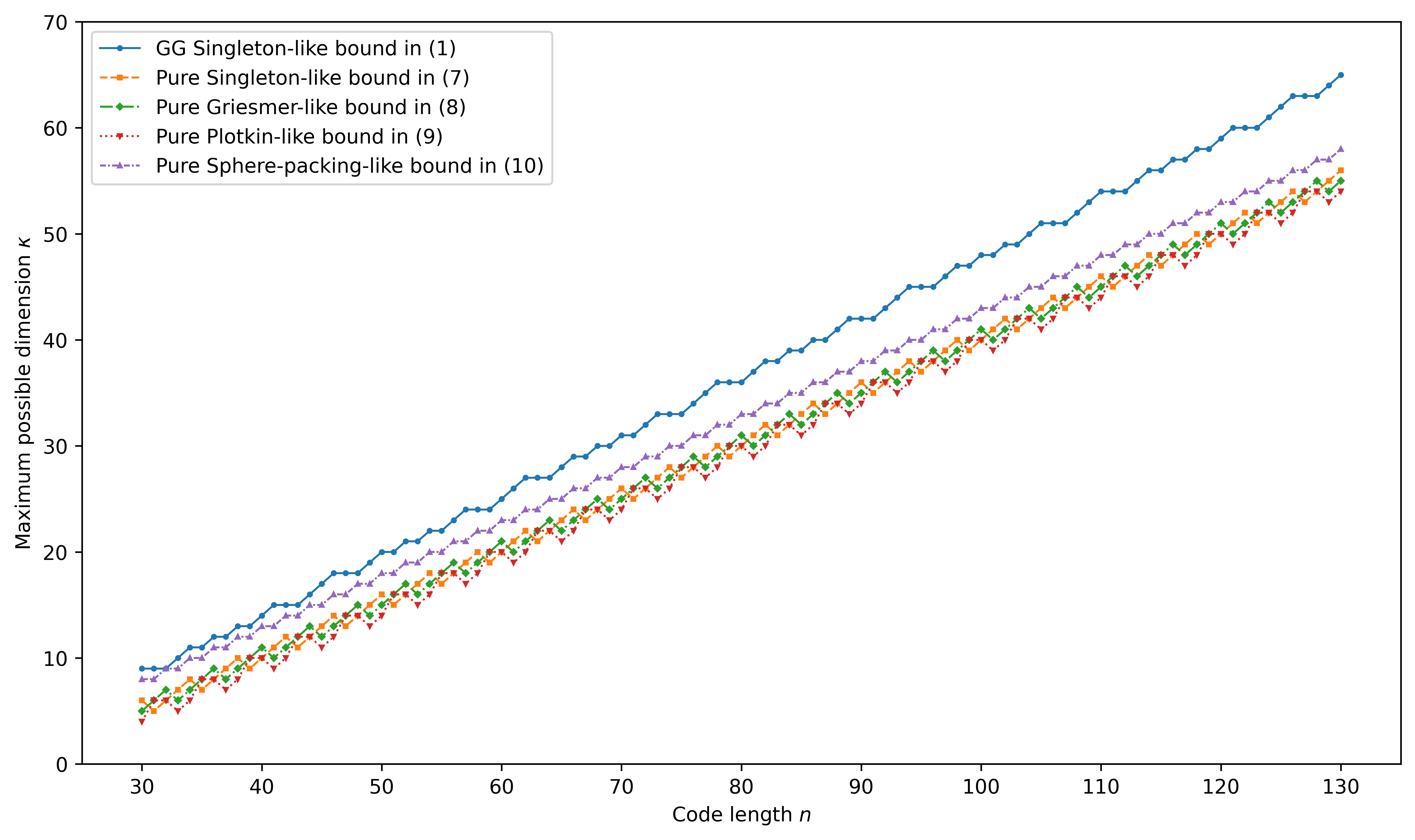} 
    \caption{Comparisons of bounds for qubits with locality $r=3$, $\delta=8$, and $30\leq n\leq 130$}
    \label{fig:bounds_finite}
\end{figure}

{\bf When the lengths are finite}, we can transform all the five bounds into upper bounds on the maximum possible dimension $\kappa$ 
such that each inequality holds for given $n$, $\delta$, $q$, and $r$ with the help of computer simulations. 
In Figure \ref{fig:bounds_finite}, we show the comparisons of these bounds for $q=2$, $r=3$, $\delta=8$, and $30\leq n\leq 130$. 
From it, one can clearly see that 
\begin{itemize}
    \item all the pure bounds in \eqref{eq.pure_Singleton}, \eqref{eq.pure_Griesmer}, \eqref{eq.pure_Plotkin}, 
and \eqref{eq.pure_Sphere-packing} are tighter than the GG Singleton-like bound in \eqref{eq.GG_Singleton}; 

    \item the pure Sphere-packing-like bound in \eqref{eq.pure_Sphere-packing} is inferior to the other three pure bounds for small $n$, 
while the other three bounds each have their pros and cons, but the pure Plotkin-like bound in \eqref{eq.pure_Plotkin} is tightest for most cases.
\end{itemize}

{\bf When the lengths are asymptotically large}, it is preferable to compare the bounds in terms of their asymptotic forms \cite{HP2003}. 
In this context, we can show that our pure bounds are always tighter than the GG Singleton-like bound in \eqref{eq.GG_Singleton} 
and reveal their tightness interrelationships, except for the pure Sphere-packing bound in \eqref{eq.Sphere-packing-like}. 
To this end, we first present the asymptotic forms of these bounds.
Given a pure $[[n,\kappa,\delta]]_q$ qLRC with locality $r$, we use  
\begin{align*}
  \mathcal{R}=\frac{\kappa}{n},~{\rm and}~\Delta=\frac{\delta}{n}
\end{align*}
to denote its {\em rate} and {\em relative distance}, respectively. 
Using the notation, the asymptotic form of the GG Singleton-like bound in \eqref{eq.GG_Singleton}  
was given in \cite[Page 1797]{LCEL2025} as follows:
\begin{align}\label{eq.GG_Singleton-like_bound_asymptotic}
  {\mathcal{R}}\leq \left(\frac{r}{r+1}\right)^2-\frac{r(2r+1)}{(r+1)^2} \Delta+o(1),~n\to \infty.
\end{align} 
Next, we present the asymptotic forms for the other four bounds. 
It follows from the pure Singleton-like bound in \eqref{eq.pure_Singleton} that 
$$
2\delta\leq n-\kappa-2\cdot \frac{n+\kappa}{2r}+4\leq \frac{r-1}{r}\cdot n-\frac{r+1}{r}\cdot \kappa +4,
$$
which implies that 
\begin{align}\label{eq.Singleton-like_bound_asymptotic}
  {\mathcal{R}}\leq \frac{r-1}{r+1}-\frac{2r}{r+1} \Delta+o(1),~n\to \infty.
\end{align}

Note that the pure Griesmer-like bound in \eqref{eq.pure_Griesmer} provides a lower bound on the code length $n$ 
given its dimension, minimum distance, and locality. Thus, it is meaningless to directly consider the limit 
$n\to\infty$ in \eqref{eq.pure_Griesmer}. To enable a fair comparison with the other three bounds, 
we instead reinterpret the pure Griesmer-like bound in \eqref{eq.pure_Griesmer} 
as an upper bound on the code size in terms of the given length, minimum distance, and locality (see also \cite{HKS2021}). 
Let $t$ be an integer such that $1\leq t\leq r$.
{By taking $\ell=\frac{n+\kappa}{2r}-1\leq \left\lceil \frac{n+\kappa}{2r} \right\rceil-1$} in the pure Griesmer-like bound in \eqref{eq.pure_Griesmer}, 
we have 
\begin{align*}
  \begin{split}
    n & \geq \left(\frac{n+\kappa}{2r}-1\right) (r+1)+\sum_{i=0}^{r-1} \left\lceil \frac{\delta}{q^{2i}} \right\rceil \\
      & \geq \left(\frac{n+\kappa}{2r}-1\right) (r+1)+\sum_{i=0}^{t-1} \frac{\delta}{q^{2i}}+(r-t) \\
      & = \frac{r+1}{2r}\cdot n + \frac{r+1}{2r}\cdot \kappa + \frac{q^{2t}-1}{q^{2t}-q^{2t-2}}\cdot \delta -(t+1).
  \end{split}
\end{align*}
which yields that 
\begin{align}\label{eq.Griesmer-like_bound_asymptotic}
  {\mathcal{R}}\leq \frac{r-1}{r+1}-\frac{2r}{r+1}\cdot \frac{q^{2t}-1}{q^{2t}-q^{2t-2}} \Delta+o(1),~n\to \infty.
\end{align}

Again, taking $\ell=\frac{n+\kappa}{2r}-1$ in the pure Plotkin-like bound in \eqref{eq.pure_Plotkin} implies that
\begin{align*}
  \delta \leq \frac{q^{2r}-q^{2r-2}}{q^{2r}-1}\cdot \left(\frac{r-1}{2r}\cdot n-\frac{r+1}{2r}\cdot \kappa +(r+1)\right),
\end{align*}
which leads to  
\begin{align}\label{eq.Plotkin-like_bound_asymptotic}
  {\mathcal{R}} \leq \frac{r-1}{r+1}-\frac{2r}{r+1}\cdot \frac{q^{2r}-1}{q^{2r}-q^{2r-2}} \Delta+o(1),~n\to \infty.
\end{align}

Since 
\begin{align}\label{eq.inequality_1}
  \frac{q^{2r}-1}{q^{2r}-q^{2r-2}}\geq \frac{q^{2t}-1}{q^{2t}-q^{2t-2}}\geq 1,~\forall~ 1\leq t\leq r, 
\end{align}
the pure Plotkin-like bound in \eqref{eq.Plotkin-like_bound_asymptotic} is tighter than the pure Griesmer-like bound in \eqref{eq.Griesmer-like_bound_asymptotic}, 
which is in turn tighter than the pure Singleton-like bound  in \eqref{eq.Singleton-like_bound_asymptotic}. 
In particular, it is easy to check that the first inequality in \eqref{eq.inequality_1} takes equality if and only if $t=r$, 
and the second inequality takes equality if and only if $t=1$. 
Therefore, if $1< t< r$, the above tighter relationship is strict, 
and such a $t$ always exists when $r\geq 3$. 
Note also that 
$$
\frac{r-1}{r+1}-\frac{2r}{r+1} \Delta < \left(\frac{r}{r+1}\right)^2-\frac{r(2r+1)}{(r+1)^2} \Delta~ \Longleftrightarrow~ \Delta\geq 0. 
$$ 
As a result, the pure Singleton-like bound in \eqref{eq.Singleton-like_bound_asymptotic} is always strictly tighter than 
the GG Singleton-like bound in \eqref{eq.GG_Singleton-like_bound_asymptotic}. 
In summary, we obtain the following results. 

\begin{theorem} 
The four asymptotic bounds in 
\eqref{eq.GG_Singleton-like_bound_asymptotic}-\eqref{eq.Plotkin-like_bound_asymptotic} 
satisfy the following hierarchy:  
\begin{align}\label{eq.hierarchy}
\begin{split}    
    &~ \text{pure Plotkin-like bound in \eqref{eq.Plotkin-like_bound_asymptotic}} \\
    \succeq &~ \text{pure Griesmer-like bound in \eqref{eq.Griesmer-like_bound_asymptotic}} \\ 
    \succeq &~ \text{pure Singleton-like bound in \eqref{eq.Singleton-like_bound_asymptotic}} \\
    \succeq &~ \text{GG Singleton-like bound in \eqref{eq.GG_Singleton-like_bound_asymptotic}},
\end{split}
\end{align}
where the notation ``$\succeq$'' indicates the relation ``is tighter than''. 
Moreover, if $r\geq 3$, then the above inequality between any 
two adjacent bounds in \eqref{eq.hierarchy} is strict.
\end{theorem}


Finally, we focus on the asymptotic form of the pure Sphere-packing-like bound in \eqref{eq.pure_Sphere-packing}. 
With this in mind, we recall the important {\em Hilbert entropy function} $\mathbf{H}_{q^2}(x)$, which is defined as 
\begin{align*}
  \mathbf{H}_{q^2}(x) = \begin{cases}
    0, & x=0, \\
    x\log_{q^2}(q^2-1)-x\log_{q^2}x-(1-x)\log_{q^2}(1-x), & 0<x\leq 1-q^{-2}.
  \end{cases}
\end{align*}
Letting 
\begin{align}\label{eq.volume_function}
    \mathbf{V}_{q^2}(n,a)=\sum_{i=0}^{a} \binom{n}{i} (q^2-1)^i,
\end{align}
we have from \cite[Lemma 2.10.3]{HP2003} that 
\begin{align}\label{eq.volume_bound}
  \frac{\log_{q^2}\mathbf{V}_{q^2}(n, \lfloor \Delta n \rfloor)}{n} =\mathbf{H}_{q^2}(\Delta), \quad n\to \infty, 
\end{align}
where $0\leq \Delta \leq 1-q^{-2}$ and $q\geq 2$. 
Let $\ell_0$ be a constant such that $0\leq \ell_0\leq \left\lfloor \frac{n-1}{r+1}\right\rfloor$. 
Using \eqref{eq.pure_Sphere-packing} and \eqref{eq.volume_function}, 
we have 
\begin{align*}
    \begin{split}
        \kappa & \leq n - 2\log_{q^2} {\mathbf{V}_{q^2}\left(n-\ell_0(r+1), \left\lfloor \frac{\Delta n-1}{2} \right\rfloor \right)}-2\ell_0 \\
        & \leq n - 2\log_{q^2} {\mathbf{V}_{q^2}\left(n-\ell_0(r+1), \left\lfloor \frac{\Delta (n-\ell_0(r+1))-1}{2} \right\rfloor \right)} -2\ell_0,   
    \end{split}
\end{align*}
which, as follows from \eqref{eq.volume_bound} and the fact that $n \to \infty$ implies $n - \ell_0(r+1) \to \infty$, yields
\begin{align}\label{eq.Sphere-packing-like_bound_asymptotic}
  {\mathcal{R}} \leq 1 - 2\mathbf{H}_{q^2}\left(\frac{\Delta}{2}\right) + o(1),~n\to \infty.
\end{align}

\begin{figure}[htbp]  
    \centering
    \includegraphics[width=0.85\textwidth]{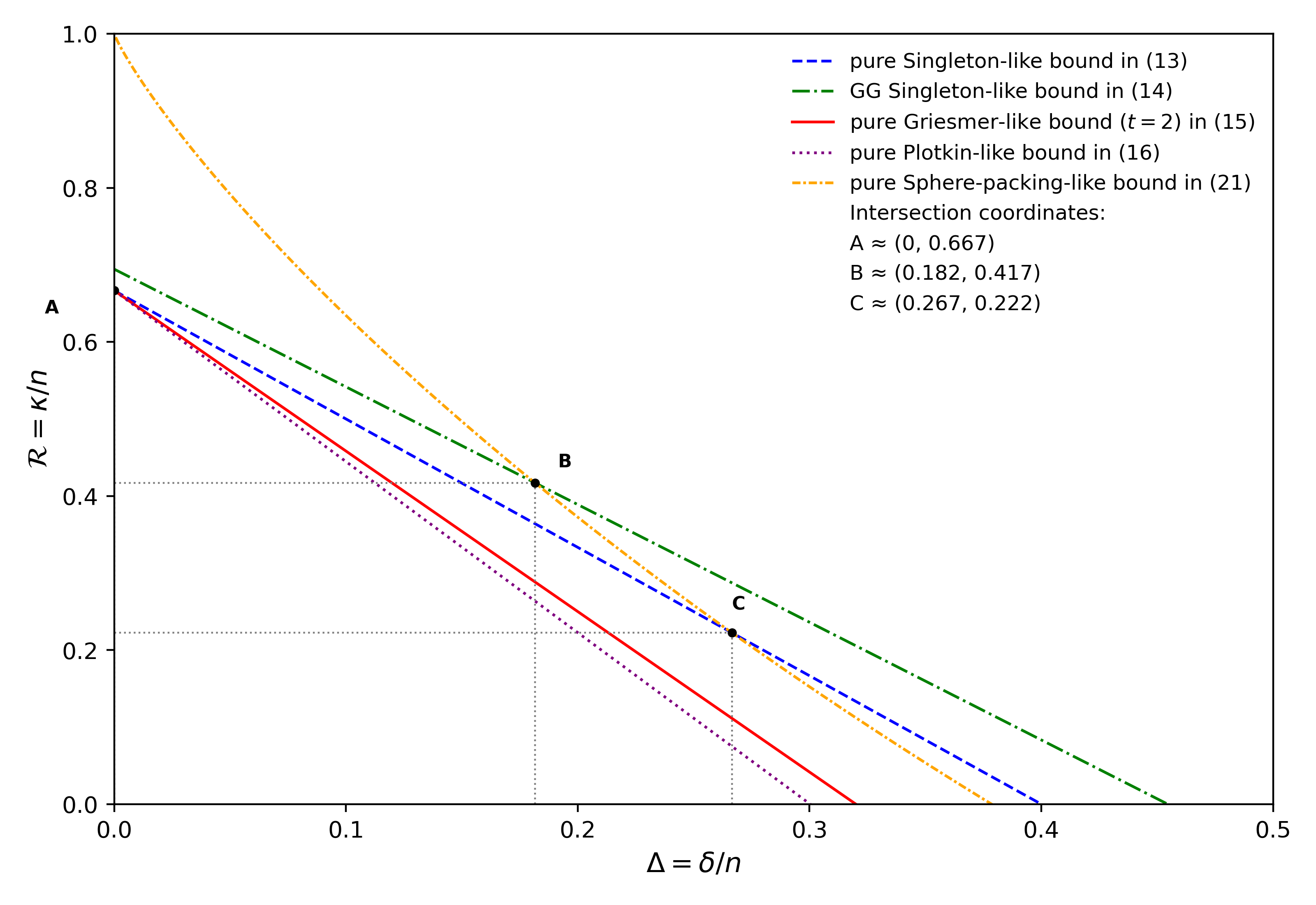} 
\caption{Comparisons of the asymptotic bounds on qubit codes with locality $5$}
\label{fig.1}
\end{figure}

\begin{figure}[htbp]  
    \centering
    \includegraphics[width=0.85\textwidth]{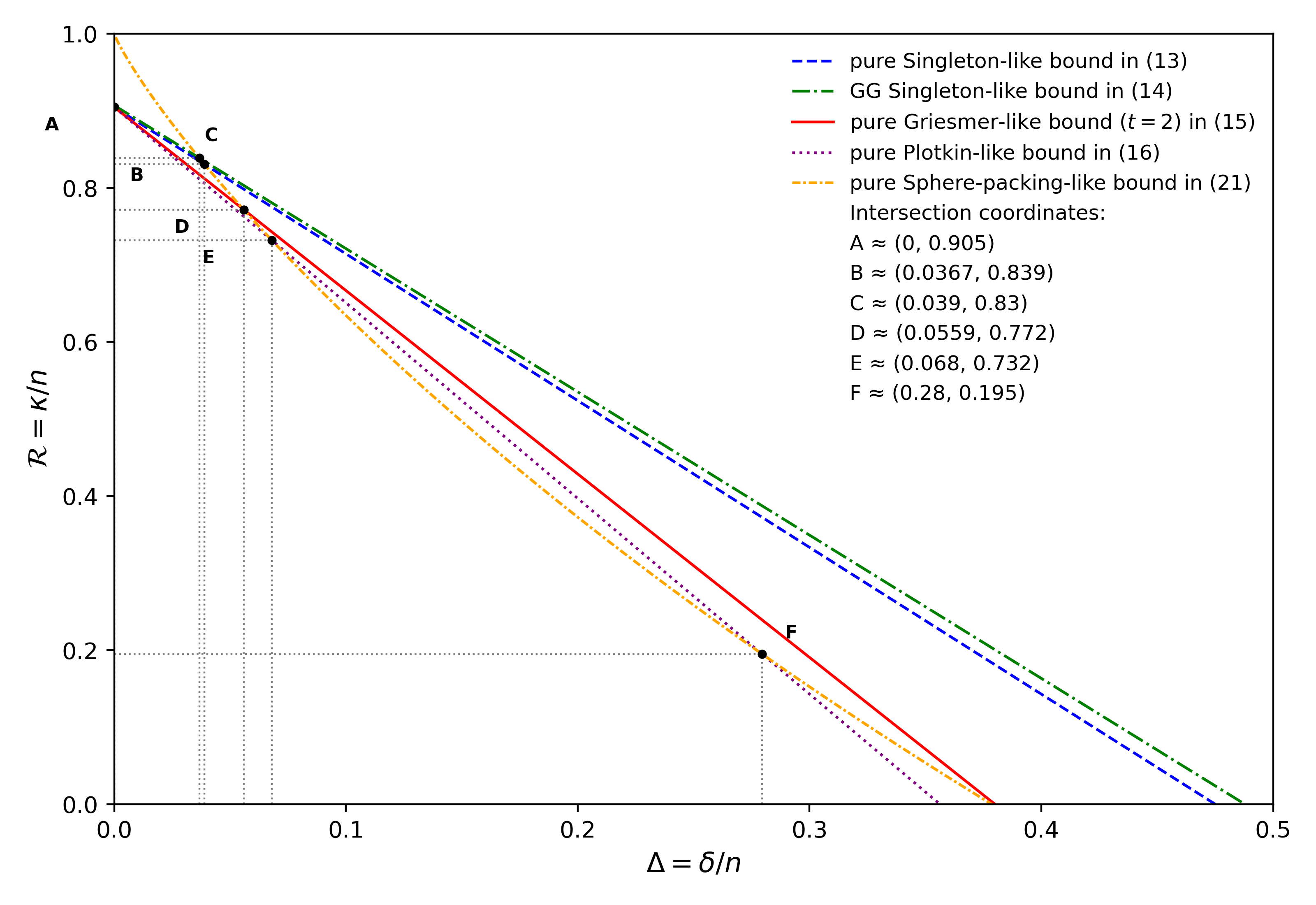} 
\caption{Comparisons of the asymptotic bounds on qubit codes with locality $20$}
\label{fig.2}
\end{figure}

However, the value of \eqref{eq.Sphere-packing-like_bound_asymptotic} is determined by the Hilbert entropy function $\mathbf{H}_{q^2}(\Delta)$, 
which is not easy to compare with the other four bounds. 
To clarify the relationship described in \eqref{eq.hierarchy} and the bound in \eqref{eq.Sphere-packing-like_bound_asymptotic},
we depict the five bounds in Figure~\ref{fig.1} for qubit codes with locality~$5$, and in Figure~\ref{fig.2} for qubit codes with locality~$20$,
where $t$ is fixed at~$2$. 
{\em These two figures not only validate the hierarchy in \eqref{eq.hierarchy},
but also reveal that \eqref{eq.Sphere-packing-like_bound_asymptotic} can be tighter than 
the other four asymptotic bounds under certain parameter regimes.} 
Note that all intersection points of these curves, along with their approximate coordinates, are also indicated in Figures~\ref{fig.1} and~\ref{fig.2}.

\section{Optimal constructions of pure qLRCs}\label{sec.4}

This section focuses on {\bf Problem} \ref{prob2}, exploring the construction of optimal pure qLRCs 
with respect to the pure bounds established in Section \ref{sec.3}.
Recall that some well-known QECCs are pure, such as the quantum Hamming codes \cite{KKKS2006}  
and quantum generalized Reed-Muller (GRM) codes \cite{SK2005}. 
Within our target, it is interesting and important to study their localities and determine whether they can form new families of 
optimal pure qLRCs with respect to these pure bounds. 
In addition, we also construct optimal pure qLRCs based on Hermitian self-orthogonal Solomon-Stiffler codes.
To this end, we need the following lemma. 

\begin{lemma}{\rm (\!\! \cite{LXY2019,TFDTZ2023})}\label{lem.locality}
    Let $\C$ be an $[n,k,d]_q$ linear code $\C$ with dual distance $d^{\perp}$. 
    For $0\leq t\leq n$, let  
\begin{align*}
    \mathcal{B}_t\left(\C\right)=\left\{\supp(\ccc):~ \ccc=(c_1,c_2,\ldots,c_n)\in \C,~ \wt(\ccc)=t\right\},   
\end{align*}
where $\supp(\ccc)=\{i: c_i\neq 0,~1\leq i\leq n\}$ and $\wt(\ccc)=|\supp(\ccc)|$. 
If $d\geq 2$ and $d^{\perp}\geq 2$, then $\C$ has locality $d^{\perp}-1$ if any one of the following conditions holds: 
\begin{enumerate}
    \item 
    $\bigcup_{\mathcal{S}\in \mathcal{B}_{d^{\perp}}\left(\C^{\perp}\right)} \mathcal{S}=\{1,2,\ldots,n\}$;
    \item 
    $\C$ is cyclic; 
    \item 
    $\C$ is affine-invariant.
\end{enumerate}

\end{lemma}

\subsection{Optimal pure quantum Hamming LRCs}

Let $H_{q^2,m}$ be a matrix over $\F_{q^2}$ whose columns are formed by selecting one nonzero vector from each one-dimensional subspace of $\F_{q^2}^m$. 
The $q^2$-ary linear code with parity-check matrix $H_{q^2,m}$ is called a {\em Hamming code} $\mathcal{H}_{q^2}(m)$.
Note that there are $q^2-1$ nonzero vectors in each one-dimensional subspace of $\F_{q^2}^m$. 
Suppose we fix a particular ordering of these one-dimensional subspaces and select one representative vector from each to 
form the parity-check matrix $H_{q^2,m}$ for $\mathcal{H}_{q^2}(m)$. 
If another parity-check matrix $H'_{q^2,m}$ is obtained by reordering the subspaces or choosing different representatives, 
then $H'_{q^2,m}$ can be derived from $H_{q^2,m}$ by rescaling and permuting its columns—equivalently, by multiplying $H_{q^2,m}$ on the right by a monomial matrix.
Hence, all such constructions define codes that are monomially equivalent.
Consequently, every Hamming code $\mathcal{H}_{q^2}(m)$ has parameters
$$\left[\frac{q^{2m}-1}{q^2-1},\frac{q^{2m}-1}{q^2-1}-m,3\right]_{q^2}$$
and its Hermitian dual is the {\em Simplex code} $\mathcal{S}_{q^2}(m)$ with parameters 
$$\left[\frac{q^{2m}-1}{q^2-1},m,q^{2m-2}\right]_{q^2}.$$

However, monomially equivalent $q^2$-ary linear codes may not preserve the Hermitian dual-containing property when $q>2$ (see, e.g., \cite{Chen2023DCC, LEGL2023}). 
It is well known that $\mathcal{H}_{q^2}(m)$ is permutation-equivalent to a cyclic code if $\gcd(m,q^2-1)=1$ and $m\geq 2$. 
In this case, $\mathcal{H}_{q^2}(m)$ was proven to be Hermitian dual-containing in \cite[Lemma 37]{KKKS2006}.
Given this, we obtain the following optimal pure qLRCs derived from $\mathcal{H}_{q^2}(m)$.

\begin{theorem}{\rm (\!\! {\bf Quantum Hamming LRCs})}\label{th.Quantum_Hamming}
    Let $m$ be an integer and let $q$ be a prime power such that $m\geq 2$ and $\gcd(m,q^2-1)=1$.  
    Then there exists a pure 
    $$\left[\left[\frac{q^{2m}-1}{q^2-1},\frac{q^{2m}-1}{q^2-1}-2m,3\right]\right]_q$$ 
    qLRC $\mathcal{Q}$ with locality $r=q^{2m-2}-1$, 
    which is optimal with respect to the pure Sphere-packing-like bound in \eqref{eq.pure_Sphere-packing}.
    Moreover, if $m\in \{2,3\}$, then $q=m^t$ with $t\geq 1$ and $\mathcal{Q}$ is also optimal 
    with respect to both the pure Singleton-like bound in \eqref{eq.pure_Singleton} and the pure Griesmer-like bound in \eqref{eq.pure_Griesmer}. 

\end{theorem}
\begin{IEEEproof}
    Since $\mathcal{H}_{q^2}(m)^{\perp_{\rm H}}\subseteq \mathcal{H}_{q^2}(m)$,
    we have $d\left(\mathcal{H}_{q^2}(m)^{\perp_{\rm H}}\right)\geq d\left(\mathcal{H}_{q^2}(m)\right)=3>2$.  
    It follows from the fact that $\mathcal{H}_{q^2}(m)$ is permutation-equivalent to a cyclic code and Lemma \ref{lem.locality}.2) 
    that $\mathcal{H}_{q^2}(m)$ has locality $r=q^{2m-2}-1$.  
    Then applying the Hermitian construction in Lemma \ref{lem.Hermitian_qLRC} to $\mathcal{H}_{q^2}(m)$ immediately yields a pure 
    \begin{align*}
      \left[\left[\frac{q^{2m}-1}{q^2-1},\frac{q^{2m}-1}{q^2-1}-2m,3\right]\right]_q
    \end{align*}
    qLRC $\mathcal{Q}$ with locality $r=q^{2m-2}-1$, 
    where the purity is derived from the inequality $q^{2m-2}>3$ for any $m\geq 2$ (see also \cite[Theorem 38]{KKKS2006}).

    Now, we discuss the optimality of $\mathcal{Q}$. 
    Note that 
    $$\left\lfloor \frac{\frac{q^{2m}-1}{q^2-1}-1}{(q^{2m-2}-1)+1} \right\rfloor 
    =\left\lfloor \frac{q^{2m}-q^2}{q^{2m}-q^{2m-2}} \right\rfloor \geq 1.$$  
    Taking $\ell=1$ in \eqref{eq.pure_Sphere-packing}, we arrive at  
    \begin{align*}
      \kappa  \leq \frac{q^{2m}-1}{q^2-1} - 2 \left( 1+ \log_{q^2} \left({1+\left(\frac{q^{2m}-1}{q^2-1}-q^{2m-2}\right)(q^2-1)} \right) \right) 
      = \frac{q^{2m}-1}{q^2-1} - 2m.
    \end{align*}
    Therefore, $\mathcal{Q}$ is optimal with respect to the pure Sphere-packing-like bound in \eqref{eq.pure_Sphere-packing}.

    Take $m=3$ as an example. Since $\gcd(3,q^2-1)=1$, we cannot have $q\equiv1\pmod3$ or $q\equiv2\pmod3$, 
    for in either of those cases $q^2\equiv1\pmod3$ and hence $3\mid(q^2-1)$, a contradiction. 
    Therefore, $q=3^t$ with $t\ge1$.
    It then follows from \eqref{eq.pure_Singleton} and taking $\ell=1$ in \eqref{eq.Griesmer-like} that 
    \begin{align*}
      2 \delta & \leq (q^4+q^2+1) - (q^4+q^2-5) -2 \cdot \left\lceil \frac{(q^4+q^2+1)+(q^4+q^2-5)}{2(q^4-1)} \right\rceil +4 = 6
    \end{align*}
    and 
    \begin{align*}
      n  \geq  (q^4-1)+1 + \sum_{t=0,~2\mid t}^{(q^4+q^2+1)+(q^4+q^2-5)-2(q^4-1)-2} \left\lceil \frac{3}{q^{t}} \right\rceil 
         = q^4 + \sum_{t=0}^{q^2-2} \left\lceil \frac{3}{q^{2t}} \right\rceil 
         = q^4+3+q^2-2=q^4+q^2+1.
    \end{align*}
As a result, $\mathcal{Q}$ is also optimal with respect to both the pure Singleton-like bound in \eqref{eq.pure_Singleton} 
and the pure Griesmer-like bound in \eqref{eq.pure_Griesmer} for $m \in \{2,3\}$. 
We have completed the whole proof. 
\end{IEEEproof}

\begin{remark}\label{rem.qLRC_from_MDS}
For $m=2$, it can be verified that the Hamming code $\mathcal{H}_{q^2}(2)$ is an MDS code. 
Consequently, the corresponding pure qLRCs described in Theorem~\ref{th.Quantum_Hamming} are in fact quantum MDS codes.
Note that Galindo $et~al.$~\cite[Proposition~37]{GHMM2024} constructed optimal pure $[[n,2k-n,n-k+1]]_q$ qLRCs 
with locality $(r,\rho)=(k,n-k+1)$ with respect to the pure Singleton-like bound, 
provided an $[n,k,n-k+1]_{q^2}$ Hermitian dual-containing MDS code exists. 
By setting $\rho=2$, their results yield optimal pure $[[n,n-2,2]]_q$ qLRCs with locality $r=n-1$. 
Clearly, Theorem~\ref{th.Quantum_Hamming} cannot be obtained directly from their results. 
Moreover, we prove that these pure qLRCs are also optimal with respect to 
the pure Griesmer-like bound in~\eqref{eq.pure_Griesmer} and 
the pure Sphere-packing-like bound in \eqref{eq.pure_Sphere-packing}. 
\end{remark}

\subsection{Optimal pure quantum GRM LRCs}

We now recall some notions on generalized Reed-Muller (GRM) codes \cite{AK1998}. 
Let $m$ be a positive integer, and let 
$(P_1,P_2,\ldots,P_n)$ be an order sequence of all the points in $\F_{q^2}^m$, so $n=q^{2m}$.
Define
\[
\F_{q^2}[x_1,x_2,\ldots,x_m]_{\leq v}=\text{span}_{\F_{q^2}}\{x_1^{i_1}x_2^{i_2}\cdots x_m^{i_m}:~ 0\leq i_1+i_2+\cdots+i_m\leq v\},
\]
the $\F_{q^2}$-vector space of all multivariate polynomials in $m$ variables of total degree at most $v$.
Let the {\em evaluation map} be
\[
\text{ev}:f\mapsto (f(P_1),f(P_2),\ldots,f(P_n))\in \F_{q^2}^n.
\]
Then the {\em generalized Reed\text{-}Muller code of order $v$} of length {$n=q^{2m}$} is defined as
\begin{align*}
  \begin{split}
    \mathcal{R}_{q^2}(v,m) & = \{\text{ev}(f)\mid f\in \F_{q^2}[x_1,x_2,\ldots,x_m]_{\leq v}\} \\
           & = \{(f(P_1),f(P_2),\ldots,f(P_n)):~ f\in \F_{q^2}[x_1,x_2,\ldots,x_m]_{\leq v}\}.
  \end{split}
\end{align*}
For $m\geq 1$ and $0\leq v\leq m(q^2-1)$, if $v$ can be written as 
$v+1=t(q^2-1)+s$ with $t\geq 0$ and $0\leq s\leq q^2-2$, then 
$\mathcal{R}_{q^2}(v,m)$ has parameters 
$$\left[q^{2m}, \sum_{i=0}^{m}(-1)^i\binom{m}{i}\binom{m+v-iq^2}{v-iq^2}, (q^2-s+1)q^{2m-2t-2}\right]_{q^2}.$$ 
Moreover, we deduce from \cite[Lemma 4]{SK2005} and the proof of \cite[Theorem 5]{SK2005} that,  
if the range of $v$ is narrowed to $0\leq v\leq m(q-1)-1$, 
then $\mathcal{R}_{q^2}(v,m)^{\perp_{\rm H}}$ is a 
$$
\left[ q^{2m}, q^{2m}-\sum_{i=0}^{m}(-1)^i\binom{m}{i}\binom{m+v-iq^2}{v-iq^2}, (s+1)q^{2t}\right]_{q^2}
$$
Hermitian dual-containing code, 
$i.e.$, $(\mathcal{R}_{q^2}(v,m)^{\perp_{\rm H}})^{\perp_{\rm H}}\subseteq \mathcal{R}_{q^2}(v,m)^{\perp_{\rm H}}$. 
In the following, we derive optimal pure qLRCs from $\mathcal{R}_{q^2}(v,m)^{\perp_{\rm H}}$.

\begin{theorem}{\rm (\!\! {\bf Quantum GRM LRCs})}\label{th.quantum_GRM}
    Let $m$, $v$ be two integers and let $q$ be a prime power such that $m\geq 1$ and $0\leq v\leq m(q-1)-1$.  
    If $v+1=t(q^2-1)+s$ with $t\geq 0$ and $0\leq s\leq q^2-2$, then there exists a pure 
    $$\left[\left[q^{2m},q^{2m}-2\sum_{i=0}^{m}(-1)^i\binom{m}{i}\binom{m+v-iq^2}{v-iq^2}, (s+1)q^{2t}\right]\right]_q$$ 
    qLRC $\mathcal{Q}$ with locality $r=(q^2-s+1)q^{2m-2t-2}-1$. 
    Moreover, the following results hold. 
    \begin{enumerate}
        \item If $v=1$ and $m\geq \max\left\{\frac{2}{q-1},1\right\}$, then $\mathcal{Q}$ is an optimal pure 
    $$\left[\left[q^{2m},q^{2m}-2m-2,3\right]\right]_q$$ 
    qLRC with locality $r=q^{2m}-q^{2m-2}-1$ 
    with respect to the pure Sphere-packing-like bound in \eqref{eq.pure_Sphere-packing}.

        \item If $v=1$ and $m = 2$, then $\mathcal{Q}$ is an optimal pure 
    $$\left[\left[q^{4},q^{4}-6,3\right]\right]_q$$ 
    qLRC with locality $r=q^4-q^2-1$ 
    with respect to both the pure Singleton-like bound in \eqref{eq.pure_Singleton} and the pure Griesmer-like bound in \eqref{eq.pure_Griesmer}.
    \end{enumerate}
\end{theorem}
\begin{IEEEproof}
    Note that GRM codes are affine-invariant and the Euclidean dual of a GRM code is still a GRM code \cite{DT2018}. 
    It then follows from Lemma \ref{lem.locality}.2) that the GRM code $\mathcal{R}_{q^2}(v,m)^{\perp_{\rm H}}$ 
    has locality $r=(q^2-s+1)q^{2m-2t-2}-1$. 
    Since $s\leq q^2-2$ and $m-t-1> 0$,
    we have ${d\left(\left(\mathcal{R}_{q^2}(v,m)^{\perp_{\rm H}}\right)^{\perp_{\rm H}}\right)}
    =d\left(\mathcal{R}_{q^2}(v,m)\right)=(q^2-s)q^{2m-2t-2}>2$.   
    Then applying the Hermitian construction in Lemma \ref{lem.Hermitian_qLRC} to $\mathcal{R}_{q^2}(v,m)^{\perp_{\rm H}}$ 
    immediately yields a pure 
    \begin{align}\label{eq.quantum_GRM_code}
        \left[\left[q^{2m},q^{2m}-2\sum_{i=0}^{m}(-1)^i\binom{m}{i}\binom{m+v-iq^2}{v-iq^2}, (s+1)q^{2t}\right]\right]_q
    \end{align}
    qLRC $\mathcal{Q}$ with locality $r=(q^2-s+1)q^{2m-2t-2}-1$, 
    where the purity is derived from \cite[Theorem 5]{SK2005}. 
    Now, we discuss the optimality of $\mathcal{Q}$ case by case. 

    1) The exact parameters of $\mathcal{Q}$ follow from substituting $v=1$ in \eqref{eq.quantum_GRM_code}. 
    For the optimality, we note that since $v=1$ and $m\geq \max\left\{\frac{2}{q-1},1\right\}$, 
    then $t=0$ and $s=2$, implying that 
    $$\left\lfloor \frac{q^{2m}-1}{((q^2-s+1)q^{2m-2t-2}-1)+1} \right\rfloor= 
    \left\lfloor \frac{q^{2m}-1}{q^{2m}-q^{2m-2}} \right\rfloor\geq 1.$$ 
    Then taking $\ell=1$ in \eqref{eq.pure_Sphere-packing}, we get that   
    \begin{align*}
      \kappa  & \leq q^{2m} -2 \left( 1+ \log_{q^2} \left({1+ q^{2m-2}(q^{2}-1)} \right) \right) \\
      & = q^{2m}-2 \log_{q^2} \left(q^{2m-2} \left(q^{2-2m}+q^2-1\right)\right)-2 \\
      & = q^{2m}-2m - 2 \log_{q^2} \left(q^{2-2m}+q^2-1\right) \\
      & < q^{2m}-2m-1,
    \end{align*}
    where the last inequality follows from $q^2-1+q^{2-2m}>q$ for any $m\geq 1$ and $q\geq 2$.
    Therefore, $\mathcal{Q}$ is optimal with respect to the pure Sphere-packing-like bound in \eqref{eq.pure_Sphere-packing}. 
    
    2) The parameters given here are obvious.   
    It then follows from \eqref{eq.pure_Singleton} and taking $\ell=1$ in \eqref{eq.Griesmer-like} that 
    \begin{align*}
      2 \delta & \leq q^4 - (q^4-6) -2 \cdot \left\lceil \frac{q^4+(q^4-6)}{2(q^4-q^2-2)} \right\rceil +4 = 6
    \end{align*}
    and 
    \begin{align*}
      n  \geq  (q^4-q^2-2)+1 + \sum_{t=0,~2\mid t}^{q^4+(q^4-6)-2(q^4-q^2-2)-2} \left\lceil \frac{3}{q^{t}} \right\rceil 
         = q^4-q^2-1 + \sum_{t=0}^{q^2-2} \left\lceil \frac{3}{q^{2t}} \right\rceil 
         = q^4-q^2-1+3+q^2-2=q^4.
    \end{align*}
    As a result, $\mathcal{Q}$ is optimal with respect to both the pure Singleton-like bound 
    in \eqref{eq.pure_Singleton} and the pure Griesmer-like bound in \eqref{eq.pure_Griesmer}.
    This completes the proof of 2).  
\end{IEEEproof}

\subsection{Optimal pure quantum Solomon-Stiffler LRCs}

In the sequel we employ Solomon–Stiffler codes \cite{SS1965} to construct new families of optimal pure qLRCs. 
It is noted that Solomon–Stiffler codes form a large family of Griesmer codes. 
Here we omit their detailed definition and properties, and focus instead on their application in constructing optimal pure qLRCs. 
Readers interested in the original constructions and further properties may consult \cite{SS1965,BM1973,H1981,H1983,H1985,HN1992,C2025}. 
In particular, {\em binary quantum Solomon–Stiffler CSS codes} were very recently constructed in \cite{XCZLL2025}.
However, no results are yet available for quantum Solomon–Stiffler codes derived via the Hermitian construction,
as the Hermitian self-orthogonality of general Solomon–Stiffler codes remains unknown.
To proceed, we will make use of the following lemma and notation. 

\begin{lemma}{\rm (\!\! \cite{LN1997})}\label{lem.power_sum}
    Let $q$ be a prime power and let $t$ be a nonnegative integer. 
    Then 
    \begin{align*}
        \sum_{x\in \F_q} x^t = \begin{cases}
            -1, & \text{if}~(q-1)\mid t~\text{and}~t> 0, \\
            0, & \text{otherwise}.
        \end{cases}
    \end{align*}
\end{lemma}

{Given an $s\times t$ matrix $A=(a_{ij})_{1\leq i\leq s,~1\leq j\leq t}$ over $\F_{q^2}$, 
we denote by $A^\dagger=(a_{ji}^q)_{1\leq j\leq t,~1\leq i\leq s}$ the {\em conjugate transpose} of $A$.} 
Then we have the following new family of pure qLRCs. 

\begin{theorem}{\rm (\!\! {\bf Quantum Solomon-Stiffler LRCs \Rmnum{1}})}\label{th.quantum_SS}
    Let $q>2$ be a prime power and let $m$, $u_1$, $u_2$, $\ldots$, $u_s$ be integers such that $m>u_1\geq u_2\geq \cdots\geq u_s\geq 2$, $\sum_{i=1}^su_i\leq m$, 
    and at most $q^2-1$ $u_i$'s take the same value. 
    If one of the following conditions holds:
    \begin{itemize}
        \item {\bf Condition A}: $\gcd(u_i,q^2-1)=1$ for any $1\leq i\leq s$, $m\geq \sum_{i=1}^su_i+2$, and $\gcd(m-\sum_{i=1}^su_i,q^2-1)=1$;
        \item {\bf Condition B}: $\gcd(u_i,q^2-1)=1$ for any $1\leq i\leq s$, and $m=\sum_{i=1}^su_i$, 
    \end{itemize}
    then there exists a pure 
 $$\left[\left[\frac{q^{2m}-1}{q^2-1}-\sum_{i=1}^s\frac{q^{2u_i}-1}{q^2-1},\frac{q^{2m}-1}{q^2-1}-\sum_{i=1}^s\frac{q^{2u_i}-1}{q^2-1}-2m,3\right]\right]_q$$
 qLRC with locality $r=q^{2m-2}-\sum_{i=1}^sq^{2u_i-2}-1$. 
\end{theorem}
\begin{IEEEproof}
    Recall that a Hamming code $\mathcal{H}_{q^2}(u_i)$ is Hermitian dual-containing whenever $u_i\geq 2$ and $\gcd(u_i, q^2 - 1) = 1$, 
    and its Hermitian dual is a Hermitian self-orthogonal Simplex code $\mathcal{S}_{q^2}(u_i)$ with parameters
    $\left[\frac{q^{2u_i} - 1}{q^2 - 1}, u_i, q^{2u_i-2}\right]_{q^2}$. 
    Note that $H_{q^2,u_i}$ is a generator matrix of $\mathcal{S}_{q^2}(u_i)$. 
    Let $$\mathbb{H}_{q^2,u_i}=\left\{{\bf h}_1,{\bf h}_2,\ldots,{\bf h}_{\frac{q^{2u_i}- 1}{q^2-1}}\right\}$$ 
    denote the set consisting of all columns of $H_{q^2,u_i}$ for $1\leq i\leq s+1$.

    Fix $u_0=0$ and $u_{s+1}=m-\sum_{i=1}^{s}u_i$. 
    Now, for $1\leq i\leq s$, let $G_i$ be a matrix whose columns consist of all distinct nonzero vectors of the form
 \[(\underbrace{0,0,\ldots,0}_{u_0+\cdots+u_{i-1}},{\bf r},{\bf r}_i)^T,\]
 where ${\bf r}^T\in \mathbb{H}_{q^2,u_i}$ and ${\bf r}_i\in \F_{q^2}^{m-\sum_{j=1}^{i}u_j}\backslash \{{\bf 0}\}$.
Similarly, let $G_{s+1}$ be a matrix whose columns consist of all distinct nonzero vectors of the form 
 \[(\underbrace{0,0,\ldots,0}_{u_1+\cdots+u_{s}},{\bf r}')^T,\]
 where ${{\bf r}'^T\in \mathbb{H}_{q^2,u_{s+1}}}$. 
 Therefore, $G_i$ is an $m\times \left( \frac{q^{2u_i}-1}{q^2-1}\cdot (q^{2m-2\sum_{j=1}^{i}u_j}-1) \right)$ matrix 
 of the form 
 \begin{align}\label{eq.G_i}
 \left(
    \begin{array}{cccc}
        O & O & \ldots & O \\
        H_{q^2,u_{i}} & H_{q^2,u_{i}} & \ldots & H_{q^2,u_{i}} \\
        \underbrace{{\bf r}_{i,1}^T~\ldots~{\bf r}_{i,1}^T}_{\frac{q^{2u_i}-1}{q^2-1}} 
        & \underbrace{{\bf r}_{i,2}^T~\ldots~{\bf r}_{i,2}^T}_{\frac{q^{2u_i}-1}{q^2-1}} & \ldots 
        & \underbrace{{\bf r}_{i,q^{2m-2\sum_{j=1}^{i}u_j}-1}^T~\ldots~{\bf r}_{i,q^{2m-2\sum_{j=1}^{i}u_j}-1}^T}_{\frac{q^{2u_i}-1}{q^2-1}}
    \end{array}
 \right)    
 \end{align}
 for $1\leq i\leq s$,
 and $G_{s+1}$ is an $m\times \frac{q^{2u_{s+1}}-1}{q^2-1}$ matrix 
 of the form 
 \begin{align}\label{eq.G_s+1}
    \left(
        \begin{array}{c}
            O \\
            H_{q^2,u_{s+1}}
        \end{array}
    \right),
 \end{align}
 where $O$ is an appropriate zero matrix.

 Denote by $\C_i$ the linear code generated by $G_i$ for any $1\leq i\leq s+1$. 
 Then $\C_i$ is Hermitian self-orthogonal if and only if $G_iG_i^\dagger=O$. 
 Since ${\bf r}_i$ runs through all nonzero vectors in $\F_{q^2}^{m-\sum_{j=1}^{i}u_j}$ for $1\leq i\leq s+1$, 
 we deduce from \eqref{eq.G_i} and $\eqref{eq.G_s+1}$ that $G_iG_i^\dagger$ vanishes if and only if the following three conditions hold:
$$
{\rm (\rmnum{1})}~ H_{q^2,u_i} H_{q^2,u_i}^\dagger=O, \quad {\rm (\rmnum{2})}~\sum_{x\in \F_{q^2}^*} x^{q}=0,\quad \text{and}, \quad {\rm (\rmnum{3})}~\sum_{x\in \F_{q^2}^*} x^{q+1}=0.
$$
On the one hand, 
Condition~${\rm (\rmnum{1})}$ holds since $\mathcal{S}_{q^2}(u_i)$ is Hermitian self-orthogonal when {\bf Condition A} or {\bf B} is given. 
On the other hand, it turns out from $q>2$ and Lemma~\ref{lem.power_sum} that Conditions~${\rm (\rmnum{2})}$ and ${\rm (\rmnum{3})}$ hold. 
Therefore, $\C_i$ is Hermitian self-orthogonal for any $1\leq i\leq s+1$.
Let
 \[G:=(G_1~|~G_2~|~\cdots~|~G_s~|~G_{s+1}).\] 
Then it can be verified that $G$ generates a Hermitian self-orthogonal code $\C$. 
According to \cite{HN1992} and \cite{SS1965}, $\C$ is further a projective Solomon-Stiffler code with dual distance $3$ and has parameters
 \[\left[n:=\frac{q^{2m}-1}{q^2-1}-\sum_{i=1}^s\frac{q^{2u_i}-1}{q^2-1},m,d:=q^{2m-2}-\sum_{i=1}^sq^{2u_i-2}\right]_{q^2}.\]
 In summary, $\C^{\perp_{\rm H}}$ is a Hermitian dual-containing code with parameters
 \[
 \left[\frac{q^{2m}-1}{q^2-1}-\sum_{i=1}^s\frac{q^{2u_i}-1}{q^2-1},\frac{q^{2m}-1}{q^2-1}-\sum_{i=1}^s\frac{q^{2u_i}-1}{q^2-1}-m,3\right]_{q^2}. 
 \]
 
 Next, we consider the locality of $\C^{{\perp_{\rm H}}}$, $i.e.,$ the locality of $\C^{{\perp}}$. 
 For $1\leq i\leq s+1$, let $\SSS_i$ be the subcode of $\C$ generated by the 
 $(\sum_{t=0}^{i-1}u_t+1)$st to $(\sum_{t=0}^{i}u_t)$th rows of $G$. 
 Then for any codeword ${\bf c}\in \C$, it can be expressed as the sum of codewords from these subcodes, $i.e.,$ 
$${\bf c}={\bf c}_1+{\bf c}_2+\cdots+{\bf c}_{s+1},$$
where ${\bf c}_i\in \SSS_i$ for any $1\leq i\leq s+1$.
By an equivalent argument in the proofs of \cite[Lemma 2.2 and Theorem 3.1]{LL2022}
\footnote{Although only the $p$-ary case with $p$ being an odd prime was considered in \cite{LL2022}, it can be checked that most of the conclusions in the paper also hold for general $q$-ary cases. Since the verification is straightforward, we will invoke this result for the $q$-ary case without providing a detailed proof.},
we conclude that ${\bf c}$ has minimum weight $d$ if and only if ${\bf c}_{i}\neq {\bf 0}$ for any $1\leq i\leq s$.
Note that for any $j\in \{1,2,\ldots,n\}$, there exists an index $1\leq i_0\leq s+1$ such that the $j$th column of $G$ is located in one certain column of $G_{i_0}$. 
 For $1\leq i\leq s+1$, since all columns of $G_{i}$ are nonzero, we can arbitrarily choose 
 ${\bf c}'_i\in \SSS_i$ such that ${\bf c}'_i\neq {\bf 0}$ 
 and ${\bf c}'_{i_0}$ has a nonzero $j$th coordinate. 
 As a result, 
 $${\bf c}'={\bf c}'_1+\cdots+{\bf c}'_{i_0-1}+\lambda {\bf c}'_{i_0}+{\bf c}'_{i_0+1}+\cdots+{\bf c}'_{s+1}$$
 forms a codeword of $\C$ with minimum weight $d$ for any $\lambda\in \F_{q^2}^*$. 
 Since the $j$th coordinate of ${\bf c}'_{i_0}$ is nonzero and $q^2>2$, there is always a $\lambda_0\in \F_{q^2}^*$ such that 
 the $j$th coordinate of ${\bf c}'$ is also nonzero. Hence, $j$ is covered by a codeword with weight $d$. 
 By using Lemma \ref{lem.locality}.1), we immediately get that $\C^{\perp_{\rm H}}$ has locality $r=d-1$.

Finally, it can be checked that 
$$
d=q^{2m-2}-\sum_{i=1}^sq^{2u_i-2}\geq q^{2m-2}-(q^2-1)\sum_{k=2}^{m-1}q^{2k-2}=q^2>3.
$$
Then applying the Hermitian construction in Lemma \ref{lem.Hermitian_qLRC} to $\C^{\perp_{\rm H}}$, 
we obtain a pure 
 \[
 \left[\left[\frac{q^{2m}-1}{q^2-1}-\sum_{i=1}^s\frac{q^{2u_i}-1}{q^2-1},\frac{q^{2m}-1}{q^2-1}-\sum_{i=1}^s\frac{q^{2u_i}-1}{q^2-1}-2m,3\right]\right]_q
 \]
 qLRC with locality $r=q^{2m-2}-\sum_{i=1}^sq^{2u_i-2}-1$. 
 This completes the proof. 
\end{IEEEproof}

In Theorem~\ref{th.quantum_SS}, we focus on the case where $q>2$.
For $q=2$, the restrictive {\bf Conditions A} and {\bf B} can be further omitted. 
Therefore, the resulting family of pure binary qLRCs admits more flexible parameters compared to the case $q>2$.

\begin{theorem}{\rm (\!\! {\bf Quantum Solomon-Stiffler LRCs \Rmnum{2}})}\label{th.quantum_SS222}
    Let $m$, $u_1$, $u_2$, $\ldots$, $u_s$ be integers such that $m>u_1\geq u_2\geq \cdots\geq u_s\geq 2$, $\sum_{i=1}^su_i\leq m$, 
    and at most three $u_i$’s take the same value.  
    Then there exists a pure 
$$\left[\left[\frac{4^{m}-1}{3}-\sum_{i=1}^s\frac{4^{u_i}-1}{3},\frac{4^{m}-1}{3}-\sum_{i=1}^s\frac{4^{u_i}-1}{3}-2m,3\right]\right]_2$$
qLRC with locality $r=4^{2m-2}-\sum_{i=1}^s4^{2u_i-2}-1$. 
\end{theorem}
\begin{IEEEproof}
  Keep the construction process in the proof of Theorem~\ref{th.quantum_SS} unchanged. 
  For $q=2$, we notice that the Hermitian self-orthogonality of quaternary Solomon–Stiffler codes has already been established in \cite[Theorem~3.15]{XLCX2024}. 
  Then by an argument similar to that in the proof of Theorem~\ref{th.quantum_SS}, we can obtain the pure qLRCs with the desired parameters. 
\end{IEEEproof}


\begin{theorem}{\rm (\!\! {\bf Quantum Solomon-Stiffler LRCs}-{\em Cont.})}\label{th.quantum_SS_optimal}
  Let $\mathcal{Q}_1$ and $\mathcal{Q}_2$ be pure quantum Solomon-Stiffler LRCs constructed in 
  Theorems~\ref{th.quantum_SS} and \ref{th.quantum_SS222}, respectively. 
  Then both $\mathcal{Q}_1$ and $\mathcal{Q}_2$ are optimal with respect to the pure Sphere-packing-like bound 
  in \eqref{eq.pure_Sphere-packing}.
\end{theorem}
\begin{IEEEproof}
  We only prove the optimality of $\mathcal{Q}_1$ and omit the proof for $\mathcal{Q}_2$ since they are similar. 
  Note that 
    $$\left\lfloor \frac{\frac{q^{2m}-1}{q^2-1}-\sum_{i=1}^s\frac{q^{2u_i}-1}{q^2-1}-1}{(q^{2m-2}-\sum_{i=1}^sq^{2u_i-2}-1)+1} \right\rfloor 
    =\left\lfloor \frac{q^{2m}-\sum_{i=1}^sq^{2u_i}+s-q^2}{q^{2m}-\sum_{i=1}^sq^{2u_i}-q^{2m-2}+\sum_{i=1}^sq^{2u_i-2}} \right\rfloor \geq 1.$$  
    Taking $\ell=1$ in \eqref{eq.pure_Sphere-packing}, we arrive at  
    \begin{align*}
      \kappa  & \leq \frac{q^{2m}-1}{q^2-1}-\sum_{i=1}^s\frac{q^{2u_i}-1}{q^2-1} -2 \left( 1+ \log_{q^2} \left({1+ \left(\frac{q^{2m}-1}{q^2-1}-\sum_{i=1}^s\frac{q^{2u_i}-1}{q^2-1}-q^{2m-2}+\sum_{i=1}^{s}q^{2u_i-2} \right)(q^{2}-1)} \right) \right) \\
      & = \frac{q^{2m}-1}{q^2-1}-\sum_{i=1}^s\frac{q^{2u_i}-1}{q^2-1}-2 \log_{q^2} \left(q^{2m-2} \left(1-\sum_{i=1}^{s}q^{2u_i-2}+sq^{2-2m}\right)\right)-2 \\
      & = \frac{q^{2m}-1}{q^2-1}-\sum_{i=1}^s\frac{q^{2u_i}-1}{q^2-1}-2m - 2 \log_{q^2} \left(1-\sum_{i=1}^{s}q^{2u_i-2}+sq^{2-2m}\right) \\
      & < \frac{q^{2m}-1}{q^2-1}-\sum_{i=1}^s\frac{q^{2u_i}-1}{q^2-1}-2m+1,
    \end{align*}
    where the last inequality follows if $1-\sum_{i=1}^{s}q^{2u_i-2}+sq^{2-2m}>q^{-1}$. 
    Moreover, such a condition always holds for the following two cases:
    \begin{itemize}
        \item If $u_s=m-1$, then $s=1$ and $u_1=m-1$. It is straightforward to verify that 
        $1-q^{2u_1-2m}+q^{2-2m}=1-q^{-2}+q^{2-2m}>q^{-1}$ for any $m> 2$ and $q> 2$;
        \item If $u_s\leq m-2$, then $1-\sum_{i=1}^{s}q^{2u_i-2}+sq^{2-2m}>q^{-1}$ is equivalent to  
        $q^{2m-2u_s-1}(q-1)+sq^2>\sum_{i=1}^{s}q^{2u_i-2u_s}$, which holds provided that 
        $q^{2m-2u_s-1}(q-1)\geq s$. Since $s\leq q^2-1$, it suffices to show that $q^{2m-2u_s-1}\geq q+1$, 
        and it is naturally true for any $m-u_s\geq 2$ and $q>2$.
    \end{itemize}
    In summary, both $\mathcal{Q}_1$ and $\mathcal{Q}_2$ are optimal with respect to the pure Sphere-packing-like bound in \eqref{eq.pure_Sphere-packing}. 
\end{IEEEproof}

\begin{remark}
  If we take $m=3$, $s=1$, and $u_1=2$ in Theorems~\ref{th.quantum_SS} and \ref{th.quantum_SS222}, 
  then $q=2^t\geq 4$ is required in Theorem~\ref{th.quantum_SS}. 
  Consequently, we obtain a unified family of pure quantum Solomon-Stiffler LRCs $\mathcal{Q}$ with parameters
  $\left[\left[q^4,q^4-6,3\right]\right]_q$ and locality $r=q^{4}-q^{2}-1$ from 
  Theorems~\ref{th.quantum_SS} and \ref{th.quantum_SS222}, where $q=2^t$ with $t\geq 2$. 
  It is worth noting that these pure quantum Solomon–Stiffler LRCs have the same parameters as those presented 
  in Theorem~\ref{th.quantum_GRM}.2), and they have been proven to be optimal with respect to both 
  the pure Singleton-like bound in \eqref{eq.pure_Singleton} and the Griesmer-like bound in \eqref{eq.pure_Griesmer}. 
  {On the other hand, Theorem~\ref{th.quantum_GRM}.2) also holds for odd $q$, 
  indicating that it cannot be obtained directly from Theorem~\ref{th.quantum_SS} or \ref{th.quantum_SS222} by simply specializing these parameters.} 
\end{remark}

We now provide an explicit example illustrating our construction.
As the structures of Hamming codes and GRM codes are well-known,
this example concentrates on optimal pure quantum Solomon–Stiffler LRCs,
with notation consistent with that used in the proofs of
Theorems~\ref{th.quantum_SS} and \ref{th.quantum_SS222}. 

\begin{example}
Let $q=2$ and let $\F_4=\{0,1,\omega,\overline{\omega}\}$, where $\overline{\omega}=\omega+1$. 
Assume that $4=m>u_2\geq u_1=2$. Then the Simplex code $\mathcal{S}_{q^2}(2)$ has a generator matrix of the form 
\[H_{4,2}=\begin{pmatrix}
10111\\
011\omega\overline{\omega}
\end{pmatrix}.\]
Let
{\setlength{\arraycolsep}{1pt}
{\small
\[G_1=\left(\begin{array}{c|c|c|c|c|c|c|c|c|c|c|c|c|c|c}
10111&10111&10111&10111&10111&10111&10111&10111&10111&10111&10111&10111&10111&10111&10111\\
011\omega\overline{\omega}&011\omega\overline{\omega}&011\omega\overline{\omega}&011\omega\overline{\omega}& 011\omega\overline{\omega}&011\omega\overline{\omega}&011\omega\overline{\omega}&011\omega\overline{\omega}& 011\omega\overline{\omega}&011\omega\overline{\omega}&011\omega\overline{\omega}&011\omega\overline{\omega}& 011\omega\overline{\omega}&011\omega\overline{\omega}&011\omega\overline{\omega}\\ \hline
00000&00000&00000&11111& 11111& 11111& 11111& \omega\omega\omega\omega\omega& \omega\omega\omega\omega\omega& \omega\omega\omega\omega\omega& \omega\omega\omega\omega\omega&\overline{\omega}\overline{\omega}\overline{\omega}\overline{\omega}\overline{\omega}&\overline{\omega}\overline{\omega}\overline{\omega}\overline{\omega}\overline{\omega}&\overline{\omega}\overline{\omega}\overline{\omega}\overline{\omega}\overline{\omega}&\overline{\omega}\overline{\omega}\overline{\omega}\overline{\omega}\overline{\omega}\\
11111& \omega\omega\omega\omega\omega&\overline{\omega}\overline{\omega}\overline{\omega}\overline{\omega}\overline{\omega} &00000& 11111& \omega\omega\omega\omega\omega&\overline{\omega}\overline{\omega}\overline{\omega}\overline{\omega}\overline{\omega}
&00000& 11111& \omega\omega\omega\omega\omega&\overline{\omega}\overline{\omega}\overline{\omega}\overline{\omega}\overline{\omega}
&00000& 11111& \omega\omega\omega\omega\omega&\overline{\omega}\overline{\omega}\overline{\omega}\overline{\omega}\overline{\omega}
\end{array}
\right).\]}}
We consider the following two cases, both of which have been verified using Magma~\cite{Magma}.
\begin{itemize}
    \item If $s=1$, then 
    \[G_2=\begin{pmatrix}
    00000\\
    00000\\
10111\\
011\omega\overline{\omega}
\end{pmatrix}\]
and it can be checked that the matrix \(G:=(G_1~|~G_2)\) generates a Hermitian self-orthogonal Solomon-Stiffer code \(\C\) 
with parameters \([80,4,60]_4\). Moreover, \(\C^{\perp_{\rm H}}\) has minimum distance \(3\) and locality \(59\).
Applying the Hermitian construction in Lemma \ref{lem.Hermitian_qLRC} to \(\C^{\perp_{\rm H}}\), 
we obtain a pure $\left[\left[80,72,3\right]\right]_2$ qLRC with locality \(59\), 
which is optimal with respect to the pure Sphere-packing-like bound in \eqref{eq.pure_Sphere-packing}. 
This result is consistent with Theorems~\ref{th.quantum_SS222} and~\ref{th.quantum_SS_optimal}.

\item If \(s=2\), then $u_2=2$ and there are no $G_2$ and $G_3$ in this case since $m=u_1+u_2$ 
implies that there are no choices for ${\bf r}_2$ and ${\bf r}'$. 
Then it can be checked that the matrix \(G:=G_1\) generates another Hermitian self-orthogonal Solomon-Stiffer code \(\C\) 
with parameters \([75,4,56]_4\). Moreover, \(\C^{\perp_{\rm H}}\) has minimum distance \(3\) and locality \(55\). 
Again, applying the Hermitian construction in Lemma \ref{lem.Hermitian_qLRC} to \(\C^{\perp_{\rm H}}\), 
we obtain a pure $\left[\left[75,67,3\right]\right]_2$ qLRC with locality \(55\), 
which is optimal with respect to the pure Sphere-packing-like bound 
  in \eqref{eq.pure_Sphere-packing}. 
This satisfies Theorems \ref{th.quantum_SS222} and \ref{th.quantum_SS_optimal}.
\end{itemize}
\end{example}

At the end of this section, we provide a comparison between our optimal pure qLRCs 
and existing results in the literature, highlighting the advantages of our constructions.

\begin{table*}[ht!]
\centering
\caption{{{Known families of optimal pure $[[n,\kappa,\delta]]_q$ qLRCs with locality $r$}}}
\label{tab:qLRCs}
\setlength{\tabcolsep}{3pt}
\renewcommand{\arraystretch}{1.3}
\resizebox{\textwidth}{!}{
\begin{threeparttable}
\begin{tabular}{c|c|c|c|c|c|l}
\toprule
No. & $[[n,\kappa,\delta]]_q$ & Maximum length  & $r$ & Condition & Optimality & Reference\\ 
\midrule

1\tnote{1} & 
$[[u(r+1),\kappa,\delta]]_{q}$ & 
$\Omega(q^2)$\tnote{2} & 
\makecell[c]{$r\leq q-1$, \\ $r>2\delta+u-4$} & 
\makecell[c]{$\kappa$ and $\delta$ are determined \\ by certain specific cases}  & 
S & 
\cite[Table \Rmnum{1}]{LCEL2025}  \\ \hline
2 & $[[q^2,q^2-6,3]]_{q}$ & $\Omega(q^2)$  & $q^2-q-1$  & $q\geq 7$ & S & \cite[Theorem 6]{XZS2025} \\\hline

3 & $[[q(q-1)/{h},q(q-1)/{h}-6,3]]_{q}$ & $\Omega(q^2/h)$  & ${q(q-h-1)}/{h}-1$ & $q\geq 7$,~$q(q-1)/h>6$ & S & \cite[Theorem 9]{XZS2025} \\ 

\midrule

4 & $\left[\left[\frac{q^{2m}-1}{q^2-1},\frac{q^{2m}-1}{q^2-1}-2m,3\right]\right]_q$ & $\Omega(q^{2m-2})$  & $q^{2m-2}-1$ & $m\geq 2$,~$\gcd(m,q^2-1)=1$ & SP  & Theorem \ref{th.Quantum_Hamming} \\\hline

5 & $\left[\left[q^2+1,q^2-3,3\right]\right]_q$ & $\Omega(q^2)$  & $q^{2}-1$ & $q=2^t$,~$t\geq 1$ & S, G, SP & Theorem \ref{th.Quantum_Hamming} \\\hline

6 & $\left[\left[q^4+q^2+1,q^4+q^2-5,3\right]\right]_q$ & $\Omega(q^4)$  & $q^{4}-1$ & $q=3^t$,~$t\geq 1$ & S, G, SP & Theorem \ref{th.Quantum_Hamming} \\\hline

7 & $\left[\left[q^{2m},q^{2m}-2m-2,3\right]\right]_q$ & $\Omega(q^{2m})$  & $q^{2m}-q^{2m-2}-1$ & $m\geq \max\left\{\frac{2}{q-1},1\right\}$ & SP & Theorem \ref{th.quantum_GRM}.1) \\\hline

8 & $\left[\left[q^{4},q^{4}-6,3\right]\right]_q$ & $\Omega(q^4)$  & $q^{4}-q^{2}-1$ & $-$ & S, G, SP & Theorem \ref{th.quantum_GRM}.2) \\\hline

9 & $\left[\left[n:=\frac{q^{2m}-1}{q^2-1}-\sum_{i=1}^s\frac{q^{2u_i}-1}{q^2-1},n-2m,3\right]\right]_q$
  & $\Omega(q^{2m-2})$
  & $q^{2m-2}-\sum_{i=1}^sq^{2u_i-2}-1$
  & \makecell[c]{$m>u_1\geq u_2\geq \cdots\geq u_s\geq 2$, \\ $\sum_{i=1}^su_i\leq m$, 
    at most $q^2-1$ \\ $u_i$’s take the same value, and the \\ 
    {\bf Condition A} or {\bf B}\tnote{3}, or $q=2$ holds}
  & SP 
  & Theorem \ref{th.quantum_SS_optimal} \\
  \bottomrule
\end{tabular}
\begin{tablenotes}
\footnotesize
\item [1] Classes 1, 2, and 3 are constructed using the CSS construction \cite{GG-QLRC2025}, and 
the Singleton-like bound used here was established for both the pure and impure qLRCs from the CSS construction in \cite{LCEL2025}.

\item [2] The maximum length of Class 1 is documented in \cite[Table \Rmnum{1}]{LCEL2025}.

\item [3] See Theorem \ref{th.quantum_SS} for the details of {\bf Conditions A} and {\bf B}.
\end{tablenotes}
\end{threeparttable}
}
\end{table*}

\begin{remark}
{\em (Comparison with known results)}
Table~\ref{tab:qLRCs} summarizes all known families of optimal pure $[[n,\kappa,\delta]]_q$ qLRCs with locality $r$,
including the new ones proposed in this paper.
Here, the abbreviations S, G, and SP denote optimality with respect to the pure Singleton-like bound in \eqref{eq.pure_Singleton},
the pure Griesmer-like bound in \eqref{eq.pure_Griesmer}, and the pure Sphere-packing-like bound in \eqref{eq.pure_Sphere-packing}, respectively.
It can be observed that the new families of optimal pure qLRCs presented in this work ($i.e.$, Classes 4–9 in Table~\ref{tab:qLRCs})
offer more flexible and diverse parameters than the previously known ones ($i.e.$, Classes 1–3 in Table~\ref{tab:qLRCs}).
In general, our optimal pure qLRCs also attain considerably larger code lengths.
\end{remark}

\section{Concluding remarks}\label{sec.conclusion}

Motivated by {\bf Problems} \ref{prob1} and \ref{prob2}, we studied improved bounds and optimal constructions of 
pure qLRCs based on the Hermitian construction. 
In Theorem~\ref{th.pure_bounds}, we established three new bounds for pure qLRCs, 
namely a pure Griesmer-like bound, a pure Plotkin-like bound, and a pure Sphere-packing-like bound. 
By means of computer-assisted evaluations for small code lengths and asymptotic analyses for sufficiently large code lengths, 
we respectively demonstrated that these new bounds are tighter than the GG Singleton-like bound in~\eqref{eq.GG_Singleton} 
and the pure Singleton-like bound in~\eqref{eq.pure_Singleton}. 
Intuitive comparisons of these bounds are illustrated in Figures~\ref{fig:bounds_finite}, \ref{fig.1}, and \ref{fig.2}.
Furthermore, we proved that some classical QECCs can be used to construct pure qLRCs, including 
quantum Hamming LRCs in Theorem \ref{th.Quantum_Hamming}, 
quantum GRM LRCs in Theorem \ref{th.quantum_GRM}, and 
quantum Solomon-Stiffler LRCs in Theorems \ref{th.quantum_SS} and \ref{th.quantum_SS222}. 
We also showed that these qLRCs contain many infinite subclasses that are optimal with respect to 
the pure Singleton-like bound in \eqref{eq.pure_Singleton}, 
the pure Griesmer-like bound in \eqref{eq.pure_Griesmer}, or 
the pure Sphere-packing-like bound in \eqref{eq.pure_Sphere-packing}. 
Finally, we summarized all known families of optimal pure qLRCs in Table~\ref{tab:qLRCs}, highlighting the advantages 
of our constructions in terms of more flexible parameters and larger code lengths.

There are also several interesting directions for future research.
\begin{enumerate}
\item Can we construct additional infinite families of optimal pure qLRCs with large code lengths and minimum distances?
The quantum GRM LRCs constructed in Theorem~\ref{th.quantum_GRM} have large code lengths, 
and their minimum distances can also be significant when suitable parameters are chosen. 
However, it remains unclear whether these subclasses are optimal with respect to the bounds developed 
in this paper or with respect to any other potential bounds.

\item Can we construct optimal pure qLRCs with respect to the pure Plotkin-like bound in~\eqref{eq.pure_Plotkin}? 
None of the known families of optimal pure qLRCs summarized in Table~\ref{tab:qLRCs} attain this bound. 
Therefore, developing new constructions of such optimal pure qLRCs would also be interesting.
\end{enumerate}



\bibliographystyle{IEEEtran}

\end{document}